\documentclass[12pt]{article}
\usepackage{amssymb,epsfig}

\begin{document}

\noindent

\def\med{{1\ov 2}}
\def\hepth#1{ {\tt hep-th/#1}}
\def\res{{\rm res}}
\def\min{{\rm min}}
\def\max{{\rm max}}
\def\inv{^{-1}}
\def\ie{{\it i.e. }}

\def\bd{\begin{description}}
\def\ed{\end{description}}
\def\etal{{\it et al. }}
\def\ie{{\it i.e. }}
\def\eg{{\it e.g. }}
\def\Cf{{\it Cf.\ }}
\def\cf{{\it cf.\ }}
\def\be{\begin{equation}}
\def\ee{\end{equation}}
\def\bes{\begin{equation*}}
\def\ees{\end{equation*}}
\def\beqa{\begin{eqnarray}}
\def\beqas{\begin{eqnarray*}}
\def\eeqa{\end{eqnarray}}
\def\eeqas{\end{eqnarray*}}
\def\bea{\begin{eqnarray}}
\def\eea{\end{eqnarray}}

\newcommand{\comps}{\mathbb{C}}
\newcommand{\reals}{\mathbb{R}}
\newcommand{\integs}{\mathbb{Z}}

\def\om{\omega}
\def\H{{\cal H}}
\def\tr{{\rm tr}}
\def\Tr{{\rm Tr}}
\def\F{{\cal F}}
\def\N{{\cal N}}

\def\d{\partial}
\def\ov{\over}

\def\pder#1#2{\frac{\partial #1}{\partial #2}}
\def\der#1#2{\frac{d #1}{d #2}}
\def\ppder#1#2#3{\frac{\partial^2 #1}{\partial #2\partial #3}}
\def\dpder#1#2{\frac{\partial^2 #1}{\partial #2 ^2 }}
\def\bemat{\left(\begin{array}}
\def\enmat{\end{array}\right)}
\def\theequation{\thesection.\arabic{equation}}
\def\e{\epsilon}
\def\cdotsh{\!\cdot}
\def\cdotsk{\!\cdot\!}

\def\Fpk{\alpha\!\cdot\!\!\F'_k}
\def\Fpu{\alpha\!\cdot\!\!\F'_1}
\def\Fpb{\beta\!\cdot\!\!\F'_1}
\def\Fpd{\alpha\!\cdot\!\!\F'_2}
\def\Fppk{\alpha\!\cdot\!\!\F''_k\!\!\cdot\! \alpha}
\def\Fppu{\alpha\!\cdot\!\!\F''_1\!\!\cdot\! \alpha}
\def\Fppb{\beta\!\cdot\!\!\F''_1\!\!\cdot\! \beta}
\def\Fppd{\alpha\!\cdot\!\!\F''_2\!\!\cdot\! \alpha}
\def\cdotsh{\!\cdot}


\begin{titlepage}
\begin{flushright}
{ ~}\vskip -1in
US-FT/22-99\\
YCTP-P32-99\\
\hepth{9911115}\\
November 1999\\
\end{flushright}

\vspace*{20pt}
\bigskip

\centerline{\LARGE ${\cal N}=2$ supersymmetric gauge theories with massive}
\bigskip
\centerline{\LARGE hypermultiplets and the Whitham hierarchy}
\vskip 0.9truecm
\centerline{\large\sc Jos\'e D. Edelstein$^{\,a}$, Marta
G\'omez--Reino$^{\,a}$,}
\vskip 0.1truecm
\centerline{\large\sc Marcos Mari\~no$^{\,b,}$\footnote{
Address after Jan. 1, 2000: Department of Physics and Astronomy, Rutgers
University, Piscataway, NJ 08855-0849, USA}  and Javier
Mas$^{\,a}$}
\vspace{1pc}

\begin{center}
{\em

$^a$ Departamento de F\'\i sica de Part\'\i culas, Universidade de Santiago de
Compostela \\
E-15706 Santiago de Compostela, Spain.\\

\bigskip
$^b$ Department of Physics, Yale University\\
New Haven, CT 06520,
USA.} \\

\vspace{5pc}

{\large \bf Abstract}

\end{center}

We embed the Seiberg--Witten solution for the low energy dynamics of ${\cal
N}=2$ super Yang--Mills theory with an even number of massive hypermultiplets
into the Whitham hierarchy. Expressions for the first and second derivatives of
the prepotential in terms of the Riemann theta function are provided which
extend previous results obtained by Gorsky, Marshakov, Mironov and Morozov.
Checks in favour of the {\em new} equations involve both their behaviour under
duality transformations and the consistency of their semiclassical expansions.

\end{titlepage}

\section{Introduction}

Soon after the paper of Seiberg and Witten \cite{SeiWitt} appeared, the
coincidence between spectral curves of soliton equations of the Toda type and
the Seiberg--Witten hyperelliptic curves for the low energy effective action of
$\N=2$ SUSY Yang--Mills was established \cite{toda,marwar}. The averaged
dynamics of these integrable systems, that goes under the generic name of
Whitham hierarchy \cite{Krichever}, allows for an interpretation of the
effective prepotential as the logarithm of a quasiclassical tau function
\cite{naktak,itomoro}. Later on, some non-perturbative results concerning the
derivatives of the prepotential as functions over the moduli space
\cite{matone,cobi} were rephrased neatly in this language \cite{eguyang,DKP}.
However, it was not until very recently that the richness of this approach
showed its true thrust. The r$\hat{\rm o}$le of the Whitham {\em times} as 
Wilsonian couplings, or infrared counterpart of microscopic deformations by
higher polynomial interactions, was well settled in the context of
two-dimensional topological conformal field theory  and the Kontsevich model
\cite{twodi}. In four dimensions, this conjectured link was put on a firm basis
thanks to the work carried out by Gorsky, Marshakov, Mironov and Morozov
\cite{ITEP}, where first and second derivatives of the prepotential with
respect to these times were computed as functions over the moduli space. The
appearance of a logarithmic derivative of the Riemann theta function confirmed
and extended analogous formulas for the contact terms in topological $\N=2$
twisted theory obtained from the so-called $u$-plane integral
\cite{mw,mm,losev,takasaki}. In Ref.\cite{emm}, the analytical results of
\cite{ITEP} were put to work. A first calculational goal was an efficient
algorithm for recursive evaluation of the semiclassical expansion of the
prepotential. Also, the relation between Whitham parameters and microscopic
deformations of pure $SU(N)$ was analysed (see also \cite{marijapon}) and,
finally, the Whitham times were seen to provide generalized spurionic sources
for breaking supersymmetry softly down to $\N=0$. In \cite{edemas}, this
formalism was used to extract (or test) non trivial strong coupling information
that is difficult to obtain from other methods as, for example, the
off-diagonal couplings at the maximal singularities of the moduli space. For a
review on the latest developments in these subjects, see
Refs.\cite{marijapon,reviews}.
\vskip2mm

In the present paper, we shall extend this formalism to the case of $\N=2$
supersymmetric Yang--Mills theory with any classical gauge group and massive
matter hypermultiplets in the fundamental representation. As it is well known,
the Seiberg--Witten ansatz also holds in this situation and one still has a
geometrical picture in terms of an auxiliary punctured Riemann surface $\Gamma$.
Also, the connection with integrable models has been observed in this case
\cite{DKP,varios}. The masses of the hypermultiplets are (linearly related to)
the residues of the Seiberg--Witten differential at the poles. From the point
of view of the theory of Riemann surfaces (and of the Whitham hierarchy), this
implies the incorporation to the game of differentials with simple poles on
$\Gamma$ (third kind meromorphic differentials). The paper is organized as
follows. We start, in Section 2, by reviewing the construction of the
universal Whitham hierarchy along the lines of Refs.\cite{Krichever,itomoro}.
In Section 3, we particularize this setup to the case of interest by
introducing the Seiberg--Witten hyperelliptic curve. Section 4 is devoted to
the construction of the generating meromorphic differential implied by the
Whitham hierarchy, in such a way that the Seiberg--Witten differential
naturally fits into the framework.
\vskip2mm

In Section 5, general formulas for the first and second derivatives of the
prepotential are obtained within this enlarged framework. They usually involve
the Riemann theta function and extend the pure gauge results of
Ref.\cite{ITEP}. In Section 6, we obtain the whole set of duality
transformations of the previously computed couplings by analyzing their
behaviour under symplectic transformations of the homology basis and
deformations of integration contours. Of course, the particular Whitham
hierarchy we are considering is strongly motivated by the Seiberg--Witten
solution for the effective $\N=2$ super Yang--Mills theory with massive
hypermultiplets whose moduli space, ultimately, should be recovered as a
submanifold. This is done in Section 7, where we end up by giving several
formulas for the first and second order derivatives of the effective
prepotential of $\N=2$ supersymmetric gauge theories. In Section 8, we give
two types of arguments supporting the expressions found in this way. First, we
show that these expressions exhibit the required duality covariance. Second,
we see by explicit computation that they are consistent in a highly
non-trivial way with the semiclassical expansion of the prepotential. In fact,
following the same lines of Ref.\cite{emm}, one can use this formalism to
develop a recursive procedure to obtain the instanton expansion of the
effective prepotential up to arbitrary order \cite{egrm}. We extend our
results to any classical gauge group in Section 9. Finally, in the last
section, we include some further remarks and present some avenues for future
research.

\section{Whitham equations and the prepotential}
\setcounter{equation}{0}

It is well known from the general theory of Riemann surfaces that there are
three basic types of Abelian differentials. They can be characterized by their
Laurent expansion about selected points called punctures. Let $P$ or $Q$
denote two such points on a Riemann surface $\Gamma$ of genus $g$, with local
coordinates $\xi_P$ and
$\xi_Q$ about them.
\bd
\item{(i)} ~{\sl Holomorphic differentials,} $d\om_i$.~ In any open set
$U\in\Gamma$, with complex coordinate $\xi$, they are of the form  $d\om =
f(\xi)
d\xi$ with $f$ an holomorphic function. The vector space of holomorphic
differentials has complex dimension $g$. If the complex curve is hyperelliptic,
\be
y^2 = \prod_{i=1}^{2(g+1)}(\lambda - e_i) ~,
\label{hyperel}
\ee
a suitable basis for these differentials is given by the following set of $g$
holomorphic 1-forms
\be
d v_k = {\lambda^{g-k} d\lambda\ov y}~,~~~k=1,2,...g ~.
\label{holom}
\ee
Given a symplectic basis of homology cycles $A^i,B_i\in H_1(\Gamma,\integs)$
one may compute their period integrals
\be
A^i{_k} = \oint_{A^i}dv_k ~~~~~~~~~~~~~ B_{ik} = \oint_{B_i}dv_k ~.
\label{periods}
\ee
A canonical basis $\{d\om_j\}$ can be defined by the $g$ linear equations
\be
\oint_{A^i} d\om_j = \delta^i{_j} ~,
\label{cicloa}
\ee
and clearly both bases are related as $d \om_j = A^{-1 k}{_j} d v_k$. Now the
matrix of $B_i$--periods yield moduli of the curve \footnote{We absorb a factor
of $(2\pi i)^{-1}$ into the definition of every integral that runs around a
closed contour so that, for example, $\oint_0 {d\xi\ov \xi} = \res_0 {d\xi\ov
\xi} = 1$.}
\be
\oint_{B_i} d\om_j = \tau_{ij} ~.
\label{ciclobe}
\ee
\item{(ii)}~ {\sl Meromorphic differentials of the second kind,}
$d\Omega^{P}_{n}$.~ These have a single pole of order $n+1$ at point
$P\in\Gamma$, and zero residue. In local coordinates $\xi_P$ about $P$,
($\xi_P(P)=0$), the normalization
\be
d\Omega^{P}_n = (\xi_P^{-n-1} + O(1))~d\xi_P ~,
\label{merseco}
\ee
determines $d\Omega^{P}_n$ up to an arbitrary combination of holomorphic
differentials $d\om_i$. To fix the regular part we shall require that it has
vanishing $A^i$--periods. Altogether, $d\Omega_n^P$ is uniquely defined by
\beqa
\res_P \xi^m_P d\Omega^P_n  &=& \delta_{mn} ~,~~~~~~~~~~\forall m > 0 ~,
\label{merpri} \\
\oint_{A^i} d\Omega^{P}_n &\rule{0mm}{8mm}=& 0 ~,~~~~~~~~~\,~~~~\forall i ~.
\label{mersecod}
\eeqa
\item{(iii)}~{\sl  Meromorphic differentials of the third kind,}
$d\Omega^{P,Q}_0$.~ These have first order poles at $P$ and $Q$ with opposite
residues taking values $+1$ and $-1$ respectively. In local coordinates
$\xi_P\,(\xi_Q)$ about $P~(Q)$
\be
d\Omega^{P,Q}_0 = (\xi_P^{-1} + {\cal O}(1))~d\xi_P = - (\xi_Q^{-1} + {\cal
O}(1))~d\xi_Q ~.
\label{mert}
\ee
We shall also normalize the regular part of these differentials by demanding
that their $A^i$--periods vanish, $\oint_{A^i}d\Omega^{P,Q}_0= 0~,~\forall i$.
\ed

Following Krichever's construction \cite{Krichever}, the moduli space of the
universal Whitham hierarchy $\hat{\cal M}_{g,p}$ is given by the set of
algebraic-geometrical data
\be
\hat{\cal M}_{g,p} \equiv \{\Gamma_g,P_a,\xi_a(P),~a=1,...,p\}
\label{moduliwh}
\ee
in which
\begin{enumerate}
\item $\Gamma_g$ denotes a smooth algebraic curve of genus $g$, and $P$ a point
on it.
\item $P_a$ are a set of $p$ points (punctures) on $\Gamma_g$ in generic
positions.
\item $\xi_a $  are local coordinates in the neighbourhood of the $p$ points,
\ie\  $\xi_a(P_a)= 0$.
\end{enumerate}

Fix a basis point $P_0$. For each given puncture $P_a, a=1,2,...,p$, a set of
{\em slow times} $T^0_{P_a,P_0}$ and $T^n_{P_a},~(n=1,2,...)$ are assigned in
correspondence with the meromorphic forms $d\Omega^{P_a,P_0}_0$ and
$d\Omega^{P_a}_n$ respectively. Defining the collective index $A=(P_a;n),~ B=
(P_b;m),~$ etc. (also including the possibility $A=(P_a,P_0;0)$), we shall write
$T^A,T^B,...$ and $d\Omega_A,~d\Omega_B,...$. In its original form, the Whitham
hierarchy can be defined  by the following set of differential equations
\be
\pder{d\Omega_A}{T^B} = \pder{d\Omega_B}{T^A} ~.
\label{Whitone}
\ee
The set of data (\ref{moduliwh}) specify the quasi-periodic integrable model
involved. For example, a Riemann surface with a single puncture provides
solutions for the KdV equation. With two singularities, solutions for the
Toda lattice can be obtained, etc.

The Whitham hierarchy may be further enhanced to incorporate also holomorphic
differentials, $d \om_i$, with associated parameters $\alpha^i$, such that the
system (\ref{Whitone}) is enlarged as follows
\be
\pder{d\om_i}{\alpha^j} =\pder{d\om_j}{\alpha^i}  ~~~;~~~
\pder{d\om_i}{T^A} = \pder{d\Omega_A}{\alpha^i} ~~~;~~~
\pder{d\Omega_A}{T^B} = \pder{d\Omega_B}{T^A}~~.
\label{Whithext}
\ee
Equations (\ref{Whithext}) are nothing but the integrability conditions
implying
the existence of a {\em generating meromorphic differential} $dS(\alpha^i,
T^A)$ satisfying
\be
\pder{~}{\alpha^i} dS = d\om_i~~~;~~~~~~ \pder{~}{T^A} dS = d\Omega_{A}~~~~.
\label{dese}
\ee
Moreover, the Whitham equations implicitely define a certain function, the {\em
prepotential}, $\F(\alpha^i,T^A)$,  through the
following set of equations
\beqa
\pder{\F}{\alpha^j} &=& \oint_{B_j} dS ~, \label{defFa} \\
\pder{\F}{T^n_{P_a}}  &=& {1\ov 2\pi i n}\res_{P_a}\,\xi^{-n}_a\, dS~ =
{1\ov 2\pi i n}\oint_{P_a} \xi^{-n} dS ~, \label{defFb} \\
\pder{\F}{T^0_{P,P_0}} &=& {1\ov 2\pi i} \int_{P_0}^P dS ~. \label{defFc}
\eeqa
The consistency of (\ref{dese})--(\ref{defFc}) follows from a direct
computation, relying solely on Riemann bilinear relations. For completeness,
the relevant information is given in Appendix A.

Due to (\ref{dese}) and the definitions in (\ref{defFa})--(\ref{defFc}), the
local behaviour of $dS$ near each puncture $P_a$ is given by
\be
dS \sim \left\{ \sum_{n\geq 1} T^{n}_{P_a} \xi_a^{-n-1} + T^{0}_{P_a,P_0}
\xi_a^{-1} + 2\pi i \sum_{n\geq 1} \, n\pder{\F}{T^{n}_{P_a}} \xi_a^{n-1}
\right\} d\xi_a ~.
\label{local}
\ee
An interesting {\em class of solutions}, and certainly that which is relevant
in connection with ${\cal N}=2$ super Yang--Mills theories, is given by those
prepotentials that are homogeneous of degree two:
\be
\sum_{i=1}^{g}\alpha^i\pder{\F}{\alpha^i} + \sum_{a=1}^p
T^0_{P_a,P_0}\pder{\F}{T^0_{P_a,P_0}} + \sum_{a=1}^p \sum_{n\geq 1}
T^n_{P_a}\pder{\F}{T^n_{P_a}} = 2\F ~.
\label{homodos}
\ee
For this kind of solutions, it is easy to see that the generating differential
$dS$ admits the following decomposition:
\be
dS=\sum_{i=1}^{g} \alpha^i d\om_i + \sum_{a=1}^p T^0_{P_a,P_0}
d\Omega^{P_a,P_0}_0 + \sum_{a=1}^p \sum_{n\geq 1} T^n_{P_a} d\Omega^{P_a}_n ~.
\label{desedos}
\ee
Indeed, it suffices to show that near each puncture $P_a$, $dS$ can be expanded
as in (\ref{local}). To this end, one finds from (\ref{dese})--(\ref{defFc})
that the local expansion of $d\om_i$ and $d\Omega^{P_a}_n$ around $P_a$
involve the second derivatives of $\F$ as follows:
\beqa
d\om_j &=& 2\pi i \sum_{m\geq 1} m \ppder{\F}{\alpha^j}{T^m_{P_a}}
\xi_a^{m-1}d\xi_a ~, \label{holo} \\
d\Omega^{P_a}_n &=& \left(\xi_a^{-n-1}+ 2\pi i\sum_{m\geq 1} m
\ppder{\F}{T^m_{P_a}}{T^n_{P_a}} \xi_a^{m-1}\right)d\xi_a ~, \label{meros} \\
d\Omega^{P_a,P_0}_0 &=&  \left(\xi_a^{-1}+ 2\pi i\sum_{m\geq 1} m
\ppder{\F}{T^m_{P_a}}{T^0_{P_a,P_0}}\xi_a^{m-1} \right) d\xi_a ~. \label{merot}
\eeqa
Inserting these expansions in (\ref{desedos}) and using (\ref{homodos}) one
arrives at (\ref{local}) as desired.

Given (\ref{desedos}) and the normalization (\ref{merpri})--(\ref{mersecod}),
we recognize that the parameters, $\alpha^i$, and the {\em slow times},
$T^n_{P_a}$, can also be recovered from $dS$,
\be
\alpha^i =  \oint_{A^i}dS ~~~,~~~ T^0_{P_a,P_0} = \res_{P_a} dS = - \res_{P_0}
dS ~~~,~~~ T^{n}_{P_a}  = \res_{P_a} \xi_a^{n} dS ~.
\label{lospara}
\ee
Finally, inserting (\ref{defFa})--(\ref{defFc}) and (\ref{lospara}) into
(\ref{homodos}), a formal expression for $\F$ in terms of $dS$ can be found,
\be
\F = \med\sum_{i=1}^{g}\oint_{A^i}dS\oint_{B_i}dS + {1\ov 4\pi i}
\sum_{a=1}^p \left( \oint_{P_a}dS\int_{P_0}^{P_a} dS + \sum_{n\geq 1}{1\ov
n}\oint_{P_a}\xi_a^{n}dS\oint_{P_a}\xi_a^{-n} dS \right) ~.
\label{prpt}
\ee

\section{The Whitham hierarchy}
\setcounter{equation}{0}

In this section we adapt the above formalism to the situation that will lead
naturally to a connection with the low-energy dynamics of asymptotically free
$\N=2$ super Yang--Mills theories with matter hypermultiplets. To this end, we
shall  specify the particular set of algebraic-geometrical data that
corresponds to the Whitham hierarchy of our interest, \ie the complex curve
$\Gamma_g$, the set of punctures and the local coordinates in their vicinities.

\subsection{The hyperelliptic curve}

Inspired by the Seiberg--Witten solution to the low-energy dynamics of
$\N=2$ super Yang--Mills theory with gauge group $SU(N_c)$ and $N_f < 2N_c$
massive hypermultiplets \cite{sunmat} we shall consider the following algebraic
curve of genus $g = N_c-1$,
\be
y^2 = (P(\lambda,u_k) + T(\lambda,m_f))^2 - 4 F(\lambda,m_f) ~,
\label{curveone}
\ee
where $P$ is the characteristic polynomial,
\be
P(\lambda;u_k) = \lambda^{N_c} - \sum_{k=2}^{N_c} u_{k} \lambda^{N_c-k} ~,
\label{charpol}
\ee
and $\beta = 2N_c - N_f$ is the coefficient of the one-loop ${\cal N}=2$ beta
function. Concerning $T$ and $F$, they are polynomials that do not depend on
the {\em moduli} $u_k$;
\be
F(\lambda,m_f) = \prod_{r=1}^{N_f}(\lambda - m_f) = \lambda^{N_f} +
\sum_{j=1}^{N_f} t_{j} \lambda^{N_f-j} ~,
\label{polefe}
\ee
and $T$ is a homogeneous polynomial in $\lambda$ and $m_f$ of degree $N_f-N_c$,
which is different from zero only when $N_f>N_c$. All dependence on $T$ can be
absorbed in a redefinition of the classical order parameters so that the
effective prepotential, the basic object of interest, does not depend
on it \cite{dhoker}. Thus, we can set $T=0$, and write the hyperelliptic curve
as follows:
\be
y^2 = P^2(\lambda,u_k) - 4 F(\lambda,m_f) ~.
\label{curve}
\ee
This curve represents a double cover of the Riemann sphere branched over $2N_c$
points. The moduli space of this genus $g=N_c-1$ curve is parametrized by the
complex numbers $u_k,~ k=2,...,N_c$. In the Seiberg--Witten model, these
complex numbers are homogeneous combinations of the vacuum expectation values
of the Casimirs of the gauge group $SU(N_c)$ and they parametrize the quantum
moduli space of vacua, while the $m_f$ are constant parameters related to the
bare masses of the hypermultiplets. As explained in the previous section, a
precise choice of the local coordinates $\xi_a$ around punctures is in order.
These coordinates are kept fixed while coordinates of moduli space are varied.
The following functions $w$ and $w\inv$,
\be
w^{\pm1}(\lambda) = \frac{P \pm y}{2\sqrt{F}} ~.
\label{ww}
\ee
provide the natural candidates to construct such well-behaved local coordinates.
In terms of them,
\be
P = {\sqrt{F}}\left(w + \frac{1}{w}\right) ~~~~~~~~~~~~~~
y = {\sqrt{F}}\left(w - \frac{1}{w}\right) ~.
\label{pyl}
\ee
 From  (\ref{ww})--(\ref{pyl}) a   relation between the variation
of the different parameters of the curve follows. Define $W \equiv P/\sqrt{F}$,
then
\be
(\d_{u_k} W) \delta {u_k} + W'\delta\lambda   + (\d_{m_f} W)\delta {m_f} =
 \frac{y}{\sqrt F}~\delta \log w ~,
\label{varicurve}
\ee
where $ (~)'$ stands for $\d_\lambda(~)$ and repeated indices are summed over.
For a given curve, that is, for fixed $u_{k}$ and $m_f$,
\be
\frac{dw}{w}= \frac{\sqrt{F}~W'}{y} d\lambda= \frac{W'}{\sqrt{W^2 -
4}}d\lambda ~.
\label{dww}
\ee
Note that these formulas are the same as for $SU(N_c)$ without matter ($N_f=0$)
\cite{ITEP}, upon replacing
\beqa
P &\longrightarrow& W \equiv P/\sqrt{F} ~~~~,~~~~~ y \longrightarrow
\tilde{y}\equiv y/\sqrt{F} = \sqrt{W^2 - 4} ~~, \nonumber\\
N_c &\longrightarrow& N\equiv N_c-1/2N_f ~, 
\label{pyn}
\eeqa
where now $W$ and $\tilde{y}$ are polinomials of order $N$ in $\lambda$.

Notice the appearance of the square root of $F$. As explained in Appendix C, at
some stage of the  computation of the derivatives of the prepotential, this
square root will be asked to be a rational function of $\lambda$. Unavoidably,
$F$ must be a square, $F=R^2$. This implies that, in the present framework,
$N_f$ must be an even integer and, moreover, massive  hypermultiplets must come
up in  degenerated pairs $m_f = m_{f+N_f/2}$. In principle one could think of
more perverse possibilities for $w^{\pm 1}$ that generalize the analogous
formulas for $SU(N_c)$. Namely, let
\be
w  = \frac{1}{2 R } (P + y)~~~~;~~~~ w^{-1}  = \frac{1}{2 R'} (P - y)
\ee
with $R  = \prod_{f=1}^{n_f}(\lambda - m'_f)~,~ R' = \prod_{f=1}^{n'_f}(\lambda
- m'_f)$, $n_f + n'_f = N_f$ and $F = RR'$. The ansatz in (\ref{ww})
corresponds to a ``symmetric scenario" where $R = R' = \sqrt{F}$ \ie $n_f =
n'_f= N_f/2$, which is the only one that preserves the involution symmetry
$(\lambda, y)\leftrightarrow (\lambda, -y) \Leftrightarrow w \leftrightarrow
w^{-1}$ between the two branches of the Riemann surface.

\subsection{Punctures and local coordinates}

Besides the curve, the set of algebraic-geometrical data demands the
specification of a set of punctures and local coordinates around them.
Again, since we are trying to embed the Seiberg--Witten solution, the
natural choice is given by the two points at infinity $(\lambda=\infty,
~y=\pm\infty)$ plus the $2N_f$ points $(\lambda=m_f, ~y=\pm P(m_f,u_k))$, that
will be denoted respectively $\infty^\pm$ and $m_f^\pm$. Following \cite{ITEP}
the local coordinates will be chosen to be the appropiate powers of $w$ that
uniformize the curve around them. More specifically, near the points
$\infty^\pm$, we have $y\sim \pm P(\lambda)\left(1 + O(W^{-2})\right)$ so that
$w^{\pm1}\sim\lambda^{N}$. Then, in the vicinities of these punctures, $\xi
\equiv w^{-1/N} \sim \lambda^{-1}$ near $\infty^+$ and $\xi \equiv w^{+1/N}
\sim\lambda^{-1}$ close to $\infty^-$. Also, near ``mass" punctures $m_f^\pm$,
the local coordinates are $\xi \equiv w^{\mp1} \sim (\lambda-m_f)$ for the
symmetric scenario.

The general framework introduced in the previous section would allow us to
consider a Whitham system (\ref{Whithext}) given by the whole set of
meromorphic differentials corresponding to these punctures. This is out of the
scope of the present paper and we will restrict ourselves to a smaller system
which is enough to accomodate the Seiberg--Witten solution of $\N=2$ super
Yang--Mills theories with massive hypermultiplets. Namely: we are not going
to include in our discussion meromorphic differentials of the second kind with
higher poles at $m_f^\pm$. According to the general prescriptions
(\ref{merseco}) and (\ref{mert}), the canonical meromorphic differentials
$d\Omega_n^{\infty^\pm}$, associated with times $T^n_{\infty^\pm}$, are
expanded as follows
\be
d\Omega_n^{\infty^\pm} \stackrel{\infty_\pm}{\longrightarrow} \left( (w^{\mp
1/N})^{-n-1}+ {\cal O}(1) \right) d(w^{\mp 1/N}) = \mp{1\ov N}\left(w^{\pm
n/N}+ {\cal O}(w^{\mp 1/N})\right) \frac{dw}{w} ~,
\label{doc}
\ee
whereas for the differentials of the third kind we have, after choosing $P_0 =
\infty^\pm~$,
\beqa
d\Omega_{0}^{m_f^\pm,\infty^\pm} &\stackrel{\infty_\pm}{\longrightarrow}& -
 \left( (w^{\mp 1/N})^{-1}+ {\cal O}(1) \right) d(w^{\mp 1/N}) = \pm{1\ov
N}\left(\frac1w+ {\cal O}(w^{\mp 1/N-1})\right) dw ~, \nonumber\\
&\stackrel{{m_f}_\pm}{\longrightarrow}& ((w^{\mp1})^{-1} + {\cal
O}(1)) ~d(w^{\mp1}) = \mp (w^{-1} + {\cal O}(w^{\mp1-1}))\, dw ~.
\label{dot}
\eeqa
They are associated with times $T^0_{m_f^\pm,\infty^\pm}$.

\section{The generating meromorphic 1-form, {\em dS}}
\setcounter{equation}{0}
\setcounter{subsubsection}{0}

In the previous sections we have reviewed the general framework of the Whitham
hierarchy and introduced the auxiliary curve. The next ingredient is the
generating 1-form $dS$. At the end of the day we will manage to identify it, on
a certain submanifold of the full Whitham moduli space, with the
Seiberg--Witten differential $dS_{SW}$ . Let us start by considering the
following set of meromorphic 1-forms
\be
d\hat{\Omega}_n \equiv \left( \left[W^{\frac{n}{N}}\right]_+ -
\frac{1}{2N_c}\sum_{f=1}^{N_f} \left.\left[W^{\frac{n}{N}}\right]_+
\right|_{\lambda=m_f}\right) \frac{dw}{w} ~,
\label{domegan}
\ee
were $[f]_+$ stands for the Laurent part of $f(\lambda)$ at $\lambda = \infty$,
\ie $[f]_+ = \sum_{k\geq 0} c_k \lambda^k$,~~($[f]_- = \sum_{k< 0} c_k
\lambda^k  = f - [f]_+$). For example,
\beqa
\left[W^{\frac{1}{N}}\right]_+ & = & \lambda - \frac{t_1}{2N} ~, \nonumber \\
\left[W^{\frac{2}{N}}\right]_+ & = &  \lambda^2 - \frac{t_1}{N}\lambda +
\frac{1+N}{2N^2} t_1^2 - \frac{t_2}{N} -\frac{2}{N}u_2 ~,\label{lad} \\
\vdots && ~~~~~~~~~~~~~~~~~~~~~\vdots \nonumber
\eeqa
Let us define for convenience the quantities $\kappa_n^f$ and $\kappa_n$,
\be
\kappa_n^f(u_k,m_s) \equiv \left.\left[W^{\frac{n}{N}}\right]_+
\right|_{\lambda=m_f} ~~~~~~~~ , ~~~~~~~~
\kappa_n(u_k,m_s) \equiv \frac{1}{2N_c} \sum_{f=1}^{N_f} \kappa_n^f ~,
\label{kappas}
\ee
in terms of which, after (\ref{dww}), the differentials $d\hat{\Omega}_n$ can
be casted in the form
\be
d\hat{\Omega}_n = \sqrt{F} \left( \left[W^{\frac{n}{N}}\right]_+ - \kappa_n
\right) W' \frac{d\lambda}{y} ~.
\label{dif}
\ee
In contrast to $d\Omega_n$, $d\hat\Omega_n$ do have non-vanishing
$A^i$--periods,
\be
c^i_n(u_k,m_f) \equiv \oint_{A^i} d\hat\Omega_n ~,
\label{aci}
\ee
as well as a non-vanishing residue at $\lambda = m_f$,
\be
b^f_n(u_k,m_s) \equiv \pm \res_{m_f^\pm} d\hat\Omega_n = \kappa_n -
\kappa_n^f ~.
\label{rmdn}
\ee
Notice that we are already working at the symmetric scenario, $\sum_{f=1}^{N_f}
b_n^f = - N~\kappa_n$ and, in particular, $b_1^f = - m_f~$.

The meromorphic differentials $d\hat\Omega_n$ are the {\em natural}
generalization of the analogous objects in pure $SU(N_c)$ \cite{ITEP}, which
can be recovered after setting, $F \to 1$ and $\kappa_n \to 0$ (as long as $N_f
\to 0$). Using (\ref{dif}), we furthermore see that $d\hat{\Omega}_1$ takes the
familiar form $d\hat{\Omega}_1 = \lambda \frac{dw}{w}$ which is the
Seiberg--Witten differential \cite{SeiWitt,sun}. Naively, one could be tempted
to consider a different set of meromorphic differentials
$d\hat\Omega_n^\prime = \left[W^{n/N}\right]_+\frac{dw}{w}~,$ which also
generalizes the pure $SU(N_c)$ case \cite{Tak}. However, the Seiberg--Witten 
differential would be in this case a combination of $d\hat\Omega_1^\prime$ and
$ d\hat\Omega_0^\prime=\frac{dw}{w}$, which in turn forces the introduction of
a new variable $T^0_{\infty^+,\infty^-}$. This extra time \footnote{The
possibility of including an extra parameter like this was earlier considered in
\cite{cobi}. It should be absent both in pure gauge and massless theories.} has
to be treated as an independent variable and introduces a number of unnecesary
complications. In particular, the derivative of $\F$ with respect to it
(\ref{defFc}) is hard to compute on general grounds. With our definition,
instead, every differential
$d\hat\Omega_n$ has  residue  balanced between $\infty$ and  all  masses $m_f$,
\beqa
\res_{\infty^\pm} d\hat\Omega_{n} & = & \res_{\infty^{\pm}} \sqrt{F}
\left(W^{\frac{n}{N}}-\left[W^{\frac{n} {N}}\right]_-\right) W'{d\lambda\ov y}
- \kappa_n\res_{\infty^{\pm}} \sqrt{F} W'{d\lambda\ov y} \nonumber\\
& = & \res_{\infty^{\pm}} \left(W^{\frac{n}{N}}\right) W'{d\lambda\ov W} \pm
N\kappa_n = \pm N\kappa_n\nonumber = - \sum_{f=1}^{N_f} \res_{m_f^\pm}
d\hat\Omega_n ~. 
\label{equalres}
\eeqa
In view of (\ref{cicloa}) and (\ref{merpri}), the above set of residues and
periods given in (\ref{aci})--(\ref{rmdn}) and (\ref{equalres}) can be taken
into account by means of the following decomposition
\be
d\hat\Omega_n = d\Omega_n + \sum_i c_n^i d\om_i + \sum_f b^f_n d\Omega_{0,f} ~,
\label{lahat}
\ee
where $d\Omega_n = -N(d\Omega^{\infty^+}_n - d\Omega^{\infty^-}_n)$, and
$d\Omega_{0,f} =d\Omega_0^{m_f^+,\infty^+} - d\Omega_0^{m_f^-,\infty^-}$.
Accordingly, $T^n$ and $T^{0,f}$ will stand for $T^n_{\infty^+} = -
T^n_{\infty^-}$ and $T^0_{m_f^+,\infty^+} = - T^0_{m_f^-,\infty^-}$
respectively. We can now proceed to evaluate the derivatives of these
differentials with respect to
$u_k$ and $m_f$ (holding $w$ fixed).

\noindent
{\bf Lemma A:}~~ The following equations hold
\beqa
\left. \pder{d\hat\Omega_n}{u_k}\right\vert_w & = & \sum_i
\left(\pder{c^i_n}{u_k}\right) d\omega_i + \sum_f
\left(\pder{b_n^f}{u_k}\right)
d\Omega_{0,f} ~, \label{va} \\
\left. \pder{d\hat\Omega_n}{m_s}\right\vert_w & = & \sum_i
\left(\pder{c^i_n}{m_s}\right) d\omega_i + \sum_f
\left(\pder{b_n^f}{m_s}\right)
d\Omega_{0,f} ~. \label{vo}
\eeqa
The proof of this lemma is given in Appendix B.

Once the basic set of meromorphic differentials has been described, we can
go one step further and set
\be
dS = \sum_{n\geq 1} T^n d\hat\Omega_n ~.
\label{dssw}
\ee
In principle this expression defines $dS(T^n,u_k,m_f)$. The idea now is
{\em to trade
the $N_c-1+N_f$   moduli  $(u_k,m_f)$, for the  equal number of Whitham
coordinates }$(\alpha^i,T^{0,f})$. In order to do so, we notice that the
structure of poles and periods of $dS$ can be taken into account in the
``integrabilistic'' basis of 1-forms, as in (\ref{desedos}):
\be
dS =  \sum_{n\geq 1} T^n d\Omega_n + \sum_{i=1}^{N_c-1}\alpha^i d\om_i +
\sum_{f=1}^{N_f} T^{0,f} d\Omega_{0,f} ~,
\label{deseint}
\ee
 Now,
$T^n$ are independent variables given by
\be
T^n =  \mp{1\ov N}\res_{\infty^+}w^{-n/N}dS ~.
\label{dow}
\ee
Using (\ref{lahat}) and (\ref{dssw}), we can compute $\alpha^i$ and $T^{0,f}$
as functions of $(T^n,u_k,m_f)$:
\beqa
\alpha^i &=&  \sum_{n\geq 1} T^n c^i_n(u_k,m_f) ~~=~~ \oint_{A^i} dS ~,
\label{dsssw} \\
T^{0,f}  &=&      \sum_{n\geq 1} T^n b^f_n(u_k,m_s)~~=~~\pm\res_{m_f^\pm}
dS ~, \label{dssww}
\eeqa
We can solve these equations for $u_k$ and $m_f$ as functions of
$(\alpha^i,T^n,T^{0,f})$, and this leads to the Whitham equations. These
equations just emphasize the r\^ole of $\alpha^i, T^{0,f}$ and
$T^n$
as independent coordinates. More explicitely, making use of (\ref{dsssw}) and
(\ref{dssww}), we  demand
\be
\frac{d\alpha^i}{ d T^n} = 0 ~ =~ c_n^i + \sum_{m\geq 1}
T^m\left(\pder{u_k}{T^n}\pder{c_m^i}{u_k} +
\pder{m_g}{T^n}\pder{c_m^i}{m_g}\right)
\label{vo2}
\ee
and
\be
\frac{d T^{0,f}}{d T^n} = 0 ~ = ~b_n^f + \sum_{m\geq 1}
T^m\left(\pder{u_k}{T^n}\pder{b_m^f}{u_k} +
\pder{m_g}{T^n}\pder{b_m^f}{m_g}\right)  ~.
\label{vo3}
\ee
In other words, calling $\rho_l= \{u_k, m_f\},~(l=1,...,N_c+N_f-1)$ the
``old'' moduli, and $\gamma_a= \{\alpha^i,T^{0,f}\},~(a=1,...,N_c+N_f-1)$ the
``new'' ones, the Whitham equations assert that $\gamma_a$ and $T^n$ form
a set of independent coordinates:
\be
\der{\gamma_a}{T^n} = \pder{\gamma_a}{T^n} +
\pder{\rho_l}{T^n}\pder{\gamma_a}{\rho_l} = 0 ~,
\label{wh1}
\ee
which can be solved for $\d u_k/\d T^n$ and $\d m_f/\d T^n$ by inverting
$\d\gamma_a/\d\rho_l~$,
\be
~~~~~~~~~~~~~~~~~~~~~~\pder{\rho_l}{T^n} = -
\pder{\gamma_a}{T^n}\pder{\rho_l}{\gamma_a} ~~ , ~~~~~~~~~~~~~ \mbox{where}
~~~~ \pder{\rho_l}{\gamma_a} \equiv
\left.\left(\pder{\gamma_a}{\rho_l}\right)^{-1}\right\vert_{T^n} ~.
\label{wh2}
\ee
The solution of these equations embodies functions $u_k$ and $m_f$ homogeneous
of degree zero in $T^n, T^{0,f}$ and $\alpha^i$. This fact follows
automatically after multiplying (\ref{wh2}) by $T^n$ and summing up in $n$.
Indeed, we see from (\ref{dsssw}) and (\ref{dssww}) that $\gamma_a$ are linear
in $T^m$, thus
\be
\sum_{n\geq 1} T^n \pder{\rho_l}{T^n} + \sum_a \gamma_a
\pder{\rho_l}{\gamma_a} = 0 ~.
\label{va2}
\ee

\noindent
{\bf Lemma B:}~~ The generating differential $dS$ satisfies
\be
\pder{dS}{\alpha^i}  = d\om_i ~~~~~~~ \pder{dS}{T^{0,f}} = d\Omega_{0,f}
~~~~~~~ \pder{dS}{T^n} = d\Omega_{n} ~.
\label{lemmab}
\ee
The proof of this lemma is left to Appendix B.

\section{Prepotential Derivatives}
\setcounter{equation}{0}
\setcounter{subsubsection}{0}

In the present section, a set of expressions is given for the dependence
of the first and second derivatives of the prepotential with respect to the
independent parameters $(\alpha^i,T^n,T^{0,f})$ as functions over the moduli
space, \ie of $u_k$ and $ m_f$. In contrast to the situation in pure super
Yang--Mills,  the parameter $N= N_c-\med N_f$ can become as small as one. For
this reason, whenever possible, we have tried to push the range of validity of
the formulas to higher times $T^n$ than those in \cite{ITEP}. In each case we
will clearly specify the allowed range.

\subsection{First Derivatives}

The formal expresions for these functions are
\beqa
\alpha^D_i ~\equiv ~~~\pder{\F}{\alpha^i} &=& \oint_{B_i} dS ~, \label{dal}\\
T^D_n ~\equiv  ~~\pder{\F}{T^n}  &=& - \frac{N}{\pi i n}
\res_{\infty^+} w^{n/N} dS ~,
\label{ccc}\\ T^D_{0,f} ~\equiv ~\pder{\F}{T^{0,f}} &=& \frac{1}{\pi i }
\int^{m_f^+}_{\infty^+} dS ~. \label{dto}
\eeqa
From these, only (\ref{ccc}) can be worked out to yield some polynomial
function of $u_k$ and $m_f$:
\beqa
\pder{\F}{T^n}  & = & - \frac{N}{\pi i n} \res_{\infty^+} w^{n/N} dS
\nonumber \\
& = & - \frac{N^2}{\pi i n^2} \sum_{m\geq 1} T^m \res_{\infty^+}
\left( \left[W^{\frac{m}{N}}\right]_+ - \kappa_m \right) dw^{n/N}. \nonumber
\eeqa
Since $w^{n/N}=W^{n/N} \left( 1- \frac{n}{N} W^{-2} + {\cal O}(W^{-4})
\right)$, this derivative takes the form
\beqa
\pder{\F}{T^n} & = & \frac{N^2}{\pi i n^2} \sum_{m\geq 1} T^m
\res_{\infty^+} W^{n/N} \left( 1
- \frac{n}{N} W^{-2} + {\cal O}(W^{-4}) \right) d\left[W^{\frac{m}{N}}\right]_+
\nonumber \\
& = & \frac{N}{\pi i n} \sum_{m\geq1} m T^m \left( {\cal H}_{n+1,m+1} -
\frac{n-2N}{N} {\cal H}_{n+1-2N,m+1} + \ldots \right) ~.
\label{dtn}
\eeqa
This expression is valid for $n+m<4N$, and the second term   gives
non-vanishing contributions only for $n+m \geq 2N$. In Eq.(\ref{dtn}), ${\cal
H}_{p+1,q+1}$ stand for polynomials in $(u_k,m_f)$ defined by
\be
{\cal H}_{p+1,q+1} = \frac{N}{pq}\res_{\infty^+} W^{\frac{p}{N}}
d\left[W^{\frac{q}{N}}\right]_+ = {\cal H}_{q+1,p+1} ~.
\label{haches}
\ee
 The first few examples of these polynomials are
\beqa
\H_{2,2} &=& u_2 -\frac{1+2N}{8N} t_1^2 + \frac{t_2}{2}, \label{h22} \\
\H_{2,3} &=& u_3 - \frac{t_1}{N} u_2 + \frac{t_3}{2} -\frac{1+N}{2N}t_1 t_2
+ \frac{1+3N+2N^2}{12N^2} t_1^3, \nonumber \\
\H_{3,3} &=& u_4 + \frac{N-2}{2N} u_2^2 + \frac{t_1^2-2t_2}{2N} u_2
+\frac{N+1}{2N}t_1^2 t_2\nonumber\\
&& ~~~ -\frac{1+N}{4N} t_2^2 - \frac{1+2N}{4N}t_1t_3 +\frac{t_4}{2}
+\frac{1-7N^2-6N^3}{48N^3}t_1^4,\nonumber\\
\vdots && ~~~~~~~~~~~~~~\vdots \nonumber
\eeqa
Concerning Eq.(\ref{dto}), let us point out that, since $dS$ has first order
poles at $m_f^\pm$, it is actually logarithmically divergent and needs a
regularization.

\subsection{Second Derivatives}

The formal expresions for these functions are obtained directly from the
general Whitham setup
adapted to the present context
\beqa
\ppder{\F}{\alpha^i}{\alpha^j} &=& \oint_{B_j} d\om_i ~, \label{daa}\\
\ppder{\cal F}{\alpha^i}{T^n} &=& -\frac{N}{\pi i n} \res_{\infty^+}
\left(w^{\frac{n}{N}} d \om_i\right) = \oint_{B_i}d\Omega_n ~, \label{dna}\\
\ppder{{\cal F}}{T^m}{T^n} &=& -\frac{N}{\pi i n} \res_{\infty^+}
\left(w^{\frac{n}{N}} d\Omega_m\right) ~,  \label{dnm}\\
\ppder{\F}{\alpha^i}{T^{0,f}} &=& \frac{1}{\pi i }\int^{m_f^+}_{\infty^+} d\om_i =
\oint_{B_i} d\Omega_{0,f} ~, \label{dor}\\
\ppder{\F}{T^{0,f}}{T^{0,g}} &=& \frac{1}{\pi i }\int^{m_f^+}_{\infty^+}
d\Omega_{0,g} ~, \label{dfg}\\
\ppder{\F}{T^{0,f}}{T^n} &=& -\frac{N}{\pi i n} \res_{\infty^+}
\left(w^{\frac{n}{N}} d \Omega_{0,f}\right) = \frac{1}{\pi
i}\int_{\infty^+}^{m_f^+}d\Omega_n ~. \label{drs}
\eeqa
For the first equation above we simply have
\be
\ppder{\F}{\alpha^i}{\alpha^j} = \tau_{ij} ~.
\label{zaa}
\ee
We shall obtain in what follows closed expressions, as functions over the
moduli space, for those derivatives involving local (residue) calculations,
such as (\ref{dna})--(\ref{dnm}) and (\ref{drs}).

\subsubsection{Mixed derivatives with respect to $T^n$ and $\alpha^i$}

The mixed derivatives with respect to $T^n$ and $\alpha^i$ (\ref{dna}) are
given by
\be
\frac{\partial^2{\cal F}}{\partial \alpha^i\partial T^{n}} = - \frac{N}{i\pi n}
{\res}_{\infty^+} w^{\frac{n}{N}} d \om_i = - \frac{N}{i\pi n}
{\res}_{\infty^+} W^{\frac{n}{N}} \left( 1 - \frac{n}{N} W^{-2} +
{\cal O}(W^{-4}) \right) d
\om_i ~.
\label{mixta}
\ee
To obtain these derivatives we still have to expand $d\om_i$ near $\infty^+$ as
follows
\beqa
d\om_i &=& \sum_k \pder{u_k}{a^i}~dv_k = \sum_k
\pder{u_k}{a^i}~\frac{\lambda^{N_c-1-k}}{P} \left( 1 + 2 W^{-2} + {\cal
O}(W^{-4}) \right)~d\lambda \nonumber \\ 
&=& - \sum_k \pder{u_k}{a^i}\pder{}{u_k}\log W
(1 + 2 W^{-2} + {\cal O}(W^{-4}))~d\lambda ~,
\eeqa
so that, finally, the residue in (\ref{mixta}) can be written as
\be
\frac{\partial^2{\cal F}}{\partial \alpha^i\partial T^{n}} = \frac{N^2}{i\pi
n^2} \frac{\partial}{\partial a_i}\res_{\infty^+} \left( W^{\frac{n}{N}} -
\frac{n-2N}{N} W^{\frac{n}{N}-2} + {\cal O}(W^{\frac{n}{N}-4}) \right)~d\lambda
~.
\label{parh}
\ee
We can better write this result in terms of the polynomials ${\cal H}_{p+1}
\equiv {\cal H}_{p+1,2} = {\cal H}_{2,p+1}$ as follows
\be
\frac{\partial^2{\cal F}}{\partial \alpha^i\partial T^{n}} = \frac{N}{\pi in}
\pder{~}{a_i} \left( {\cal H}_{n+1} - \frac{n-2N}{N}{\cal H}_{n-2N+1}
+ ... \right) ~,
\label{dan}
\ee
where the dots denote terms that contribute only for $n\geq 4N-1$, and the
derivative $\d/\d a^i$ should be taken at constant $m_f$.

\subsubsection{Second derivatives with respect to $T^n$ and $T^m$}

The second derivatives of the prepotential with respect to the Whitham times
are given by
\be
\frac{\partial^2{\cal F}}{\partial T^{m}\partial T^{n}} = - \frac{N}{i\pi n}
{\res}_{\infty^+} w^{\frac{n}{N}} d \Omega_{m} = \frac{N^2}{i\pi n}
{\res}_{\infty^+} w^{\frac{n}{N}} \left(d\Omega_{m}^{\infty^+} -
d\Omega_{m}^{\infty^-} \right) ~.
\label{cc}
\ee
To evaluate this residue, it is more convenient to use the canonical
differentials in hyperelliptic coordinates
$d\tilde{\Omega}_m^{\infty^\pm}(\lambda)$. The relation between
$d\Omega_m^{\infty^\pm}$ and $d\tilde\Omega_m^{\infty^\pm}$ is easily obtained
by matching the asymptotic behaviour around $\infty^\pm$. Expanding $W^{m/N} =
\sum_{p=-\infty}^m b_{mp}^{(N)}\lambda^p$ one gets (for $m\leq$ 2N)
$$
d\Omega_m^{\infty^\pm} \stackrel{(\ref{doc})}{=} -\frac{1}{m}dw^{\pm
m/N} + \cdots = \frac
1m\sum_{p=1}^mpb_{mp}^{(N)}(-\lambda^{p-1}d\lambda+\cdots)=  \frac
1m\sum_{p=1}^mpb_{mp}^{(N)}d\tilde\Omega_p^\pm ~,
$$
where the dots denote the regular part of the differentials, which is
unambiguously fixed by the condition (\ref{mersecod}). Eq.(\ref{cc}) can now be
written as
\be 
\frac{\partial^2{\cal F}}{\partial T^{m}\partial T^{n}} = \frac{N^2}{i\pi
nm}\sum_{p=1}^{m}p \,b_{mp}^{(N)} \res_{\infty^+} \left(w^{\frac{n}{N}} d
\tilde{\Omega}_{p}\right) ~. 
\label{interm}
\ee
This residue can be computed with the help of expression (\ref{c12}), and
closely follows the one performed in \cite{ITEP} for the pure gauge
theory. It turns out that, as long as $m+n<2N$, the result found there is still
valid (with the obvious replacements given in (\ref{pyn}))
\be
\frac{\partial^2{\cal F}}{\partial T^{m}\partial T^{n}} = - \frac{N}{\pi
i} \left( {\cal H}_{n+1,m+1} + \frac{2N}{mn} \frac{\partial{\cal
H}_{n+1}}{\partial a_i}\frac{\partial {\cal H}_{m+1}}{\partial
a_j}\frac{1}{i\pi}\partial_{\tau_{ij}}
\log\Theta_E  \right) ~,
\label{ssasd}
\ee
where $\Theta_E$ is the Riemann theta function with a particular even and
half-integer characteristic (see Appendix C). In asymptotically free theories
with paired massive hypermultiplets, $N$ can be as small as $1$. Thus, we
would need to extended the range of validity for this formula to higher times
$T^n,T^m$ up to $1\leq n,m \leq 2N$ with $n+m\leq 2N$, in order to
include at least $T^1$. In general, when $n+m\geq 2N$, additional
contributions must be considered. If $n+m=2N$, it is just a constant
\be
\frac{N}{i\pi} \left( \frac{m}{n} + \theta(m-N-1)\frac{4(N-m)N}{mn}
\right)\delta_{m+n,2N} = \frac{N}{i\pi}\,f_{mn}\delta_{m+n,2N} ~,
\label{aldasd}
\ee
where $f_{mn} = \min(m,n)/\max(m,n)$. Then, the net result is
\be
\frac{\partial^2{\cal F}}{\partial T^{m}\partial T^{n}}  = - \frac{N}{\pi i}
\left( {\cal H}_{n+1,m+1} + \frac{2N}{mn} \frac{\partial {\cal
H}_{n+1}}{\partial a_i}\frac{\partial {\cal H}_{m+1}}{\partial
a_j}\frac{1}{i\pi}\partial_{\tau_{ij}}\log\Theta_E - f_{nm}\delta_{n+m,2N}
\right) ~.
\label{ss}
\ee

\subsubsection{Second derivatives with respect to $T^n$ and $T^{0,f}$}

The calculation of the second derivative of the prepotencial $\F$ with respect
to $T^n$ and $T^{0,f}$ requires the evaluation of the residue
\be
\ppder{\F}{T^{0,f}}{T^n} = -\frac{N}{\pi i n} \res_{\infty^+} w^{n/N}
d\Omega_{0,f} =-\frac{N}{\pi i n} \res_{\infty^+} w^{n/N}
 \left(d\Omega_{0}^{m_f^+,\infty^+}-
d\Omega_{0}^{m_f^-,\infty^-}\right).
 \label{deri}
\ee
Like in the preceding derivation, calculations are feasible in hyperelliptic
coordinates $\lambda$. In this case, however, it is clear that
$d\tilde\Omega_{0}^{m_f^\pm,\infty^\pm}(\lambda) =
d\Omega_{0}^{m_f^\pm,\infty^\pm}(w(\lambda))$ since the coefficient of
both singular parts, \ie the residue, is fixed to be
$\pm 1$. Remember also that we are working in the ``symmetric
scenario'' in which the masses come degenerated in pairs; the index $f$ will
then run in the range $f=1,\cdots,N_f/2$. It is possible to obtain, from the
general theory of Riemann surfaces \cite{fay,rauch}, a convenient
representation for the meromorphic differential of the third kind
$d\Omega_{0,f}$ (the details of this calculation are given in Appendix C)
\be
d \Omega^{m_f^+,\infty^+}_{0} - d \Omega^{m_f^-,\infty^-}_{0} =
\frac{P}{y}\frac{d\lambda}{(\lambda-m_f)}+\frac{1}{\pi i} d\omega_i~\partial_i
\log\Theta_E(\vec z_f \mid \tau) ~,
\ee
where the vector $\vec z_f$,
\be
\vec z_f =\displaystyle{{1\ov 2\pi i}\int^{m_f^+}_{\infty^+}d \vec \om} ~,
\label{abelpoint}
\ee
is the image of the divisor ${m_f^+}-{\infty^+}$ under the Abel map. Inserting
the previous formula in (\ref{deri}), we obtain
\be
\ppder{\F}{T^{0,f}}{T^n} = ~\frac{N}{\pi in}\kappa_n^f
~+~\frac{N}{\pi^2 n}\pder{H_{n+1}}{a^i}~\partial_i
\log\Theta_E(\vec z_f \mid \tau) ~,
\label{dod}
\ee
this result being valid as long as $n\leq 2N$.

\section{Duality transformations}
\setcounter{equation}{0}

As it is well-known, one of the key properties of the Seiberg--Witten ansatz is
the existence of equivalent duality frames for the low-energy theory. In the
theories with matter hypermultiplets, as already remarked in \cite{SeiWitt},
the duality transformations (which are usually elements of the symplectic
group) pick an inhomogeneous part associated with the masses of the
hypermultiplets. We will show in what follows how all this  is nicely encoded
in the present geometrical framework. The duality symmetry will turn out to be
nothing but an ambiguity in the choice of the geometrical data involved in the
construction of the prepotential within the Whitham hierarchy. We start
by characterizing this ambiguity. For this purpose it is convenient to recall
equation (\ref{prpt}). As it stands, it is just a formal expression but it
nicely exhibits the fact that the different duality frames are associated with
different choices of integration contours. We will distinguish two types of
operations that can be performed on these contours:
\bd
\item{-} {\em changes of the symplectic homology basis}  $(A^i,B_j)\to (
\tilde A^i, \tilde B_j)$. These are performed as usual by means of a matrix
$\Gamma$,
\be
\Gamma = \bemat{cc} A & B \\ C & D \enmat \in {\rm Sp}(2r,{\mathbb Z}) ~,
\label{sympl}
\ee
where $r$ is the rank of the gauge group and the $r \times r$ matrices $A$,
$B$, $C$, $D$ satisfy:
\be
A^t D-C^t B = {\bf 1}, \,\,\,\,\,\ A^tC= C^t A, \,\,\,\,\,\ B^tD= D^t B.
\label{conds}
\ee
\item{-} {\em deformations of the integration contours}.  When we deform a
one-cycle across a pole, we pick up the residue of $dS$. In principle, these
deformations can be performed independently on each integration contour.
\ed
Bearing in mind the equations (\ref{dow})--(\ref{dssww}) and
(\ref{dal})--(\ref{dto}), we see that the previous operations leave $T^n,
T^D_n$ and $T^{0,f}$ intact
\be
\tilde T^n = T^n ~, ~~~~~~~~~ \tilde T^{D}_n = T^D_n ~, ~~~~~~~~~~
\tilde T^{0,f} = T^{0,f} ~.
\label{trtn}
\ee
Therefore, the most general ambiguity yields the ansatz
\beqa
\tilde\alpha^D_i &=& A_i^{~j} \alpha^D_j + B_{ij}\alpha^j + p_{i
f}T^{0,f} ~, \label{tra1} \\
\tilde\alpha^{i} &=&  C^{ij} \alpha^D_j + D^i_{~j}\alpha^j + q_f^i
T^{0,f} ~, \label{tra2} \\
\tilde T^{D}_{0,f} & = & T^D_{0\,f} + R_{fi} \,\alpha^i + S^i_f\,\alpha^D_i
+ t_{fg}\, T^{0,g} ~, \label{trans}
\eeqa
together with (\ref{trtn}), where $T^{0,f}$ appears on the right hand side
because it is the residue of $dS$ at the pole $\lambda=m_f$. $R_{fi}$ and
$S^i_f$ are matrices of integers that signal the possibility of deforming the
contour between $m_f$ and $\infty$ by encircling additional cycles. Also,
$p_{i f}$ and $q_f^i$ are even (because of the paired masses) integer
coefficients that account for poles that are crossed when one-cycle
deformations are performed. The extended duality group is, however, not as big
as these formulas may suggest. Namely, the deformations in the contours that
define
$T^D_{0,f}$,
$\alpha^i$ and
$\alpha^D_i$ {\em cannot occur independently}. This is not easy to see
geometrically, but it is a consequence of the fact that a single function, the
prepotential, is behind the whole construction.  The  representation  given in
(\ref{prpt}) is nothing but  the statement of the fact that  $\F$ is a
homogeneous function of degree two. For the present case it can be written as
\be
\F = \frac{1}{2} \left( \alpha^i\alpha^D_i + T^n T^D_n + T^{0,f} T^D_{0,f}
\right) ~,
\label{hom}
\ee
where repeated indices $i,n$ and $f$ are summed up. Plugging
(\ref{trtn})--(\ref{trans}) in this formula, and using (\ref{conds}), one
easily obtains the transformed $\tilde\F$. This expression can be used to
compute $\d \tilde\F/ \d \alpha^i$, which should be compared with
\be
\pder{\tilde \F}{\alpha^i} =
\pder{\tilde \alpha^j}{\alpha^i}\,\tilde\alpha^D_j ~.
\label{consist}
\ee
Agreement between both expressions enforces the constraint:
\be
S^i_f = p_{j f}C^{ij} - q^j_f D^i{_j}~~~~~~;~~~~~~~
R_{i f} = p_{j f} A_i{^j} -  q^j_f  B_{ij}~.
\ee
Hence (\ref{trans}) reduces to
\be
\tilde T^{D}_{0,f} = T^D_{0,f} + p_{i f}\bigl( C^{ij}\alpha^D_j +
D^i_{~j} \alpha^j) - q_f^i(A_i^{~j} \alpha^D_j + B_{ij} \alpha^j) + t_{fg}
T^{0,g} ~,
\label{dualtr}
\ee
and, from Eq.(\ref{hom}), we obtain the generalized duality transformation rule
for $\F$ (see also \cite{luisIyII}):
\bea
\tilde {\cal F}(\tilde \alpha^i,T^n,T^{0,f}) & = & {\cal
F}(\alpha^i,T^n,T^{0,f}) + {1 \over 2} \alpha^i (D^TB)_{ij}\alpha^j + {1 \over
2} \alpha^D_i (C^T A)^{ij} \alpha^D_j + \alpha^D_i (B^T C)^i_{~j}
\alpha^j \nonumber \\ 
& + & p_{i f}T^{0,f} (C^{ij} \alpha^D_j + D^i_{~j} \alpha^j)  + \med (p^i_f
q_{ig} + t_{fg}) T^{0,f}T^{0,g} ~.
\label{genFt}
\eea
Using this expression, one can easily compute the transformation properties
of the first and second derivatives of the prepotential. Reserving the indices
$i,j,k$ for the cycles $\alpha^i$ and $\alpha^D_i$, ~$f,g$ for the times
associated to the masses $T^{0,f},T^D_{0,f}~$, and $n,m$ for the higher Whitham
times
$T^n, T^D_n$, we define the generalized couplings
\beqa
\nonumber
\tau_{ij}= {\partial^2 {\cal F}\over \partial \alpha^i\partial \alpha^j} ~~~~~~
\tau_{if}= {\partial^2 {\cal F}\over \partial \alpha^i \partial T^{0,f}} ~~~~~~
\tau_{in}= {\partial^2 {\cal F}\over \partial \alpha^i \partial T^n} ~~~~~~ \\
\tau_{fg}= {\partial^2 {\cal F}\over \partial T^{0,f} \partial T^{0,g}} ~~~~~~
\tau_{fn}= {\partial^2 {\cal F}\over \partial T^{0,f} \partial T^n } ~~~~~~
\tau_{nm}= {\partial^2 {\cal F}\over \partial T^n T^m} ~.
\label{coupl}
\eeqa
To unravel the transformation rules for these couplings the most efficient way
is to make use of their geometrical definition (\ref{daa})--(\ref{drs}).
The geometrical data involved are contours, residues and differentials.
After the previous discussions, the contours change as follows
\beqa
\oint_{\tilde B_i} &=& A_i{^j}\oint_{B_j} + ~B_{ij}\oint_{A^j} +~ \sum_f\,
p_{i f}\,\res_{m_f} ~, \label{cont1} \\
\oint_{\tilde A^i} &=& C^{ij}\oint_{B_j} + ~D^i{_j}\oint_{A^j} +~ \sum_f\,
q^i_f \,\res_{m_f} ~, \label{cont2} \\
 \widetilde{\int_{m_f}^\infty} &=& \int_{m_f}^\infty + ~p_{if}
(C^{ij}\oint_{B_j} + D^i{_j}\oint_{A^j}) - ~q^i_f (A_i{^j}\oint_{B_j} +
B_{ij}\oint_{A^j}) + ~ \sum_g  \,t_{fg}\,\res_{m_g} ~. \label{cont3}
\eeqa
The change in the symplectic homology basis can be easily pulled back to the
canonically normalized basis of meromorphic differentials
\beqa
d\tilde\om_i &=& d\om_j (C\tau + D)^{-1~j}{_i} ~, \label{difs1} \\
d\tilde\Omega_n &=& d\Omega_n - d\om_i~((C\tau + D)^{-1}C)^{ij}
\oint_{B_j}d\Omega_n ~, \label{difs2}\\
d\tilde\Omega_{0,f} &=&  d\Omega_{0,f} - d\om_i~((C\tau + D)^{-1}C)^{ij}
\oint_{B_j}d\Omega_{0,f}) - (C\tau + D)^{-1~i}{_j} q^j_f ~. \label{difs3}
\eeqa
Inserting (\ref{cont1})--(\ref{cont3}) and (\ref{difs1})--(\ref{difs3}) in
(\ref{daa})--(\ref{drs}), the transformation rules for the couplings
(\ref{coupl}) come out straightforwardly
\beqa
\tilde\tau_{ij} & = & \left[(A\tau + B)(C\tau+ D)^{-1}\right]_{ij} ~,
\label{tauij}
\\ {\tilde\tau}_{im} & = & \left[ (C\tau + D)^{-1}\right]^j{_i} {}~\tau_{jm} ~,
\label{tauim} \\
\tilde\tau_{mn} & = & \tau_{mn} - \tau_{im}\left[(C\tau + D)^{-1} C
\right]^{ij} \tau_{jn} ~, \label{taumn} \\
\tilde\tau_{if} &=&  \left[ (C\tau + D)^{-1}\right]^j{_i} {}~\tau_{jf} -
\left[(A\tau + B)(C\tau+ D)^{-1}\right]_{ij}q_f^j+p_{i f} ~, \label{tauif} \\
\tilde\tau_{fn} &=& \tau_{fn} - q_f^i\left[(C\tau + D)^{-1} \right]^j{_i}~
\tau_{jn} - \tau_{if}\left[(C\tau + D)^{-1} C \right]^{ij} \tau_{jn} ~,
\label{taufn} \\
\tilde\tau_{fg} &=& \tau_{fg} - \tau_{if}\left[(C\tau + D)^{-1} C
\right]^{ij}\tau_{jg} + q^i_f \left[(A\tau + B)(C\tau +
D)^{-1}\right]_{ij}q^j_g
\nonumber\\ &-& q_f^i \left[ (C\tau + D)^{-1}\right]^j{_i}~ \tau_{jg} - q_g^i
\left[ (C\tau + D)^{-1}\right]^j{_i}~ \tau_{jf}  - p_{if}q^i_g + t_{fg} ~.
\label{taufg}
\eeqa
Eventually, we find another contraint on $t_{fg}$ from the requirement of
symmetry under $f\leftrightarrow g$ in the last expression. This is solved in
general by taking $t_{fg} =  p^{~}_{i[f}{q^i}_{g]} + s_{(fg)}$ with $s_{(fg)}$
an arbitrary integer valued symmetric matrix. It is reassuring to find that
(\ref{genFt}) and (\ref{tauij})--(\ref{taufg}) fully coincide and generalize
the results presented in \cite{emm} for pure $SU(N_c)$ and \cite{luisIyII}, to
which they reduce when there is only one higher Whitham time $T^n\sim
\Lambda\delta_{n1}$.

\section{The Seiberg--Witten hyperplane}
\setcounter{equation}{0}

In this section we shall identify the Seiberg--Witten solution as a submanifold
of the Whitham configuration space. In the former, the $a^i$ variables of the
prepotential, for the duality frame associated to the $A^i$--cycles, are
given by the integrals over these cycles of a certain meromorphic one-form,
$dS_{SW}$, that can be written as
\be
a^i(u_{k}, m_f;\Lambda) = \oint_{A^i} dS_{SW} \equiv \oint_{A^i} {\lambda
W'(u_{k}, m_f)\ov \sqrt{W^2(u_{k}, m_f) -4 \Lambda^{2N}}}d\lambda ~,
\label{laa}
\ee
and the same expression holds for the dual variables $a^D_i$ with $B_i$
replacing $A^i$. Here, $\Lambda$ stands for $\Lambda_{N_f}$, the quantum
generated dynamical scale. On the Whitham side, correspondingly, we have the
$\alpha^i$ variables given by
\bea
\alpha^i(u_{k},m_f; T^n) &=& \oint_{A^i}dS= \sum_{n\geq 1} T^n
\oint_{A^i}d\hat\Omega_{n} \nonumber\\
&=& \sum_{n\geq 1} T^n \oint_{A^i}{ (\left[W(u_{k},m_f)^{\frac{n}{N}}\right]_+
- \kappa_n) W'(u_k,m_f)\ov \sqrt{W^2(u_k,m_f) - 4} }  d\lambda ~,
\label{lus}
\eea
and the same for their duals $\alpha^D_i$ changing $A^i$ by $B_i$. Using
(\ref{laa}), one easily sees that
\beqa
\alpha^i(u_{k},m_f; T^1, T^2,...) &=& T^1 a^i(u_{k},m_f;1) + {\cal O}(T^{n>
1}) ~\nonumber\\
&=& a^i(\bar u_{k},\bar m_f,\Lambda=T^1) +{\cal O}(T^{n> 1}) ~,
\label{alpha}
\eeqa
where we have introduced the set of rescaled variables \footnote{Notice that
the variable $a^i$ in (\ref{dan}), (\ref{ss}) and (\ref{dod}) stands precisely
for $a^i(u_{k},m_f;1)$.}
\be
\bar u_{k} = (T^1)^{k} ~u_{k} ~~~~~,~~~~~ \bar m_{f} = T^1~ m_f ~.
\label{cv}
\ee
Summarizing, the Seiberg--Witten differential of \cite{sunmat} can be exactly
recovered after performing dimensional analysis in units of the scale set by
$T^1$, and tuning $T^{n>1} = 0$. Using (\ref{dssww}), we also see that the
Whitham times $T^{0,f}$ become (up to a sign) the bare masses:
\be
T^{0,f} = - \bar m_f ~~~~~~~~~~~ f = 1, ..., N_f/2 ~.
\ee
In view of the previous considerations, we shall define the following change of
variables $(\alpha, T^{0,f},T^{n>1}, T^1)\to (\alpha,  T^{0,f},\bar T^{n>1},
\log\Lambda)$, where
\be
\log\Lambda=\log T^1 ~,~~~~~~~ \bar T^n=(T^1)^{-n} T^n ~~~~(n>1) ~,
\label{cambio}
\ee
and, consequently,
\be
\pder{~}{\log\Lambda} = \sum_{m\geq 1} m T^m \pder{~}{T^m} ~, ~~~~~~~~
\pder{~}{\bar T^n} =  (T^1)^n \pder{~}{T^n} ~~~~(n>1) ~. 
\label{cambio2}
\ee
With the help of these expressions, one can rewrite all the formulas given in
the last section for the derivatives of $\F$, as derivatives with respect to
$\alpha^i, T^{0,f}$, $\bar T^{n>1}$ and $\log \Lambda$. Most of the factors
$T^1$ are used to promote $u_k$ to $\bar u_k$ or, rather, the homogeneous
combinations thereof
\be
\bar \H_{m+1,n+1} = (T^1)^{m+n}\H_{m+1,n+1} ~~~~ \Rightarrow ~~~~
\bar \H_{m+1} = (T^1)^{m+1}\H_{m+1} ~,
\label{hachesbar}
\ee
the remaining ones are absorbed in making up $\bar a^i\equiv T^1
a^i(u_k,m_f;1) = a^i(\bar u_k,\bar m_f; T^1)$ and $\bar m_f\equiv T^1m_f$ (see
\cite{emm} for this explicit intermediate step).

At the end of the day, the restriction to the submanifold $\bar T^{n>1}=0$ and
$T^1=\Lambda$, yields formulas which  are ready for use in the Seiberg--Witten
analysis. Notice that in this subspace $\alpha^i(u_k,m_f; T^1=\Lambda,T^{n> 1} =
0)=a^i(\bar u_k,\bar m_f,\Lambda=T^1)\equiv \bar a^i $; hence (after omitting
all bars for clarity) one can write
\be
\pder{\F}{\log \Lambda\,} = {N\ov \pi i} \left( \H_2 + \Lambda^{2}\delta_{N,1}
\right) ~~~, ~~~ \ppder{\F}{a^i}{\log \Lambda} ~=~ {N\ov \pi i} \pder{\H_{2}}{
a^i} ~, ~~~
\label{lasecpri} 
\ee
\be
\dpder{\F}{(\log\Lambda)} = - {2N^2\ov \pi i} \pder{\H_2}{ a^i} \pder{\H_2}
{ a^j}\, {1\ov \pi i} \d_{\tau_{ij}}\log\Theta_E(0|\tau)
+ \frac{2N^2}{\pi i} \Lambda^2 \delta_{N,1} ~,
\label{lasecu} 
\ee
\be
\ppder{\F}{m_f}{\log \Lambda} = - \frac{N}{\pi i}~\left( m_f-\frac{
t_1}{2N}\right) - \frac{N}{\pi^2 } ~ \pder{\H_{2}}{ a^i}  ~\d_i
 \log\Theta_E(\vec z_f \mid \tau) ~,
\label{lasecudeg}
\ee
as well as, for derivatives with respect to higher Whitham times, we obtain
\be
\pder{\F}{ T^n} = {N\ov \pi i n} \left( \H_{n+1} +
\frac{1}{N}\Lambda^{2N}\delta_{n,2N-1} \right) ~~~, ~~~ \ppder{\F}{a^i}{T^n}
~=~ {N\ov \pi i n}\pder{\H_{n+1}}{ a^i} ~, 
\label{lagun} 
\ee
\be
\ppder{\F}{\log\Lambda\,}{ T^n} = - {2N^2\ov \pi in}\pder{\H_{n+1}}{ a^i}
\pder{\H_{2}}{ a^j}\, {1\ov \pi i} \d_{\tau_{ij}}\log\Theta_E(0|\tau)
+ \frac{n+N}{\pi in} \Lambda^{2N} \delta_{n,2N-1} ~,
\label{lasecudef} 
\ee
\be
\ppder{\F}{m_f}{T^n} = - \frac{N}{\pi in} \kappa_n^f -
\frac{N}{\pi^2 n} ~ \pder{ \H_{n+1}}{a^i}  ~\d_i
 \log\Theta_E(\vec z_f \mid \tau) ~,
\label{lasfn}
\ee
\beqa
\ppder{\F}{T^{m}}{ T^n} &=& - {N\ov \pi i} \left( \H_{n+1,m+1}+{2N\ov mn}
\pder{\H_{n+1}}{a^i}\pder{\H_{m+1}}{ a^j}{1\ov \pi i}
\d_{\tau_{ij}}\log\Theta_E(0|\tau) \right. \nonumber \\
&& \left. - \frac{\min(m,n)}{\max(m,n)} ~\Lambda^{2N} \delta_{n+m,2N}
\right)  ~,
\label{lasemn}
\eeqa
where $t_1=-2\sum_{f=1}^{N_f/2} m_f$ (\cf Eq.(\ref{polefe})) and
$m,n\geq 2$, whereas $n \leq 2N$ and $m+n \leq 2N$ in (\ref{lasemn}). It is
worth to remark that, whereas the latter set of equations
(\ref{lagun})--(\ref{lasemn}) involve deformations of the effective
prepotential parametrized by higher Whitham times, the former one
(\ref{lasecpri})--(\ref{lasecudeg}) is entirely written in terms of the
original Seiberg--Witten variables.

Finally, one can combine Eqs.(\ref{lagun})--(\ref{lasfn}) to write the
following interesting expressions for the derivatives of (homogeneous
combinations of) higher Casimir operators with respect to $\Lambda$ and $m_f$,
\be
\pder{\H_{n+1}}{\log\Lambda\,} = - 2N\pder{\H_{n+1}}{ a^i}\pder{\H_{2}}{ a^j}
{1\ov \pi i} \d_{\tau_{ij}}\log\Theta_E(0|\tau) +
\frac{N-1}{N}\Lambda^{2N}\delta_{n,2N-1} ~,
\label{higheruno}
\ee
\be
\pder{\H_{n+1}}{m_f\,} = - \kappa_n^f + \frac{1}{\pi i} ~
\pder{\H_{n+1}}{a^i}  ~\d_i \log\Theta_E(\vec z_f \mid \tau) ~.
\label{higherdos}
\ee
Let us provide in the following section some non-trivial checks supporting
these results.

\section{Some checks}
\setcounter{equation}{0}

One of the main results in this paper is given by the whole set of equations
(\ref{lasecpri})--(\ref{higherdos}) for the derivatives of the effective
prepotential. The equation (\ref{lasecu}) is by now a well settled result. In
Ref.\cite{losev}, an independent derivation coming from topological field
theory was obtained prior to the work \cite{ITEP}. For the pure gauge $SU(2)$
case, it was checked by using the Picard--Fuchs equations in \cite{ITEP}.
In Refs.\cite{emm,egrm}, it was put in the test bench and two additional proofs
were passed. First of all, the right hand side was shown to reproduce correctly
the appropriate duality transformation rules. In addition, this equation was
used to obtain the semiclassical expansion of the effective prepotential up to
arbitrary instanton corrections with remarkable sucess. In this section, we
will see that equation (\ref{lasecudef}) also enjoys the generalized duality
properties and is consistent with the instanton expansion.

Let us first analyze the duality transformations (\ref{trtn})--(\ref{tra2}) and
(\ref{dualtr}) where the new ingredient, as we have already remarked, is the
inhomogeneous piece associated to the presence of masses. This inhomogeneous
piece stems from the deformation of contours across simple poles
(\ref{cont1})--(\ref{cont3}). We can reinterpret this ambiguity in the context
of the formulas involving the Riemann Theta function. First of all, the vector
$\vec z_f$ (\ref{abelpoint}) lives in the Jacobian of the hyperelliptic curve,
as the image of the divisor ${m_f^+}-{\infty^+}$ under the Abel map, thus
being defined up to transformations of the form:
\be
z_{f,i} \rightarrow z_{f,i} + n_{f,i} + \tau_{ij}\ell^{j}_f
\label{inho}
\ee
with integers $n_{f,i}$ and $\ell^{j}_f$. Taking into account that $z_{f,i} =
1/2 ~\tau_{if}$ (\cf \ref{dor}), we see that (\ref{inho}) reproduces the
formula (\ref{tauif}), when the symplectic rotation is the identity, with
$p_{i,f} = 2n_{i,f}$, $q^j_f = -2\ell^j_f$. Now we can check that the formula
(\ref{dod}) is consistent with the transformation law given in
equation (\ref{taufn}). Using the transformation property of the theta function
under shifts
$$
\Theta \left[ {\vec\alpha\atop \vec\beta}\right] (z_{f,i} + n_{f,i} +
\tau_{ij}\ell^j_f|\tau) = \exp \bigl( -\pi i \ell^i_f\tau_{ij}\ell^j_f - 2\pi i
\ell^i_f (z_{f,i} + \beta_i) + 2\pi i \alpha^i n_{f,i} \bigr)\Theta \left[
{\vec\alpha\atop \vec\beta}\right](z_{f,i}|\tau) ~,
$$
we easily obtain
\be
{1 \over \pi i } {\partial\over \partial z_i} \log \Theta_E (\vec z_f|\tau)
\rightarrow - 2\ell^i_f + {1 \over \pi i } {\partial\over \partial z_i} \log
\Theta_E (\vec z_f|\tau) ~,
\label{shth}
\ee
which induces precisely the transformation law (\ref{taufn}) with $q^i_f =
-2\ell^i_f$. Next, for the behavior of the Theta function under homogeneous
symplectic transformations we find
\be
{1 \over 2 \pi i} {\partial \over \partial z_i} \log
\Theta_E(\vec z_f|\tau) \rightarrow {1 \over 2 \pi i}
{\partial \over \partial z_i} \log \Theta_E(\vec z_f|\tau) +
\left[(C\tau + D)^{-1} C \right]^{ij}z_{f,j} ~,
\label{homogsym}
\ee
and taking into account (\ref{dan}), we find again the right transformation
properties in the whole set of expressions (\ref{tauij})--(\ref{taufg}).

A second, and much more stringent check of equation (\ref{lasecudeg}) is
provided by the semiclassical expansion of the prepotential in powers of
$\Lambda$. One can write the  ansatz
\bea
\F & = & \frac{3N}{4N_c\pi i}\sum_{\alpha_+} Z_{\alpha_+}^2
+ {i\ov 4 \pi} \sum_{\alpha_+} Z_{\alpha_+}^2 \log \,{Z_{\alpha_+}^2\ov
\Lambda^2}-{i\ov 4\pi}
\sum_{p=1}^{N_c}\sum_{f=1}^{N_f/2}(e_p+m_f)^2\log\,{{(e_p+m_f)^2}\ov
{\Lambda^2}} \nonumber\\
&& -{N_f\ov 8\pi i}\sum_{f=1}^{N_f/2}m_f^2\log\,{{m_f^2}\ov
{\Lambda^2}}+{t_1^2\ov 16\pi i}\log\,{{t_1^2}\ov{\Lambda^2}} +
{1\ov 2 \pi i}\sum_{k=1}^\infty\F_{k}(Z)\Lambda^{2kN} ~.
\label{elprep}
\eea
The set $\{ \alpha_i \}_{i=1,...,r}$ stands for the simple roots of the
corresponding classical Lie algebra. Also in (\ref{elprep}), $\alpha_+$ denotes
a positive root and $\sum_{\alpha_+}$ is the sum over all positive roots.
The dot product $(\cdot)$ of two simple roots $\alpha_i$ and $\alpha_j$ gives
an element of the Cartan matrix, $A_{ij}=\alpha_i\cdotsk\alpha_j$
and extends bilinearly to arbitrary linear combinations of simple roots.
For any root  $\alpha = n^j\alpha_j\in\Delta$, the quantities $Z_\alpha$ are
defined  by $Z_\alpha = a\cdotsk \alpha \equiv a^i A_{ij} n^j$ where $a=
a^i\alpha_i$. Simple roots can be written in terms of the orthogonal set of
unit vectors $\{\e_p\}_{p=1,\cdots,N_c}$ and the order parameters $a^i$ and
$e_p$ are related by $e_p = a\cdotsk\e_p$. Also $\F_k(a^i, m_f)$ are
homogeneous functions of degree $2- 2kN$ that represent the instanton
corrections to the perturbative 1-loop effective action.

In Refs.\cite{emm,egrm}, it was shown that inserting (\ref{elprep}) into
(\ref{lasecu}), and expanding both members of the equation in powers of
$\Lambda^{2N}$, the instanton corrections $\F_k$ could be fixed completely in
a recursive way. As an example, for $SU(2)$ with two degenerated flavors ($N_f
= 2$) one readily obtains
\beqa
\F_1 &=& \rule{0mm}{10mm}\frac{u_2+m^2}{2u_2} ~,
\label{su2f1}\\
\F_2 &=& \rule{0mm}{10mm}\frac{u^2_2-6u_2 m^2 + 5 m^4}{64 u^3_2} ~,
\label{su2f2} \\
\F_3 &=& \rule{0mm}{10mm}\frac{5u^2_2 m^2-14u_2 m^4+9m^6}{192u_2^5} ~...
\label{su2f3}
\eeqa
where $u_2$ stands for the quadratic polynomial $u_2 = a^2$. Once the
prepotential expansion has been solved up to a certain power of $\Lambda$, it
can be inserted in expression (\ref{lasecudeg}). Matching both sides is  
highly non-trivial since this equation involves simultaneously three different
types of couplings, namely $\tau_{f\Lambda}$  on the left hand side, and
$\tau_{if}$ and $\tau_{ij}$ as arguments of the Riemann theta function on the
right hand side. We have checked on the computer that, indeed, this equation is
satisfied order by order for $SU(2)$ with $N_f = 2$ up to $\Lambda^{6}$, for
$SU(3)$ with $N_f=2$ and $4$ up to $\Lambda^{8}$ and $\Lambda^4$ respectively,
and for $SU(4)$ with $N_f = 2$ and $4$ up to order $\Lambda^{12}$ and
$\Lambda^8$. We believe that this test gives a strong support to the
expression (\ref{lasecudeg}).

Note in passing that, as compared to the usual ansatz for $\F$, the fourth and
fifth terms in (\ref{elprep}) have been added for consistency of all the
equations. These terms do  not depend  on  $a_i$ and, being linear in
$\log\Lambda$, they only contribute  to the derivatives
$\pder{\F}{\log\Lambda}$, $\ppder{\F}{T^{0,f}}{\log\Lambda}$ and
$\ppder{\F}{T^{0,f}}{T^{0,g}}$; neither to the couplings nor to the
instanton expansion. So, they correspond to a freedom of the prepotential that
is fixed by the embedding into the Whitham hierarchy. A similar feature was
observed before in the uses of this framework to study the strong coupling
regime of ${\cal N}=2$ pure gauge theories near the maximal singularities of
the quantum moduli space \cite{edemas}.

\section{Other Gauge Groups}
\setcounter{equation}{0}

We can extend the results of previous sections to all classical gauge groups
$SO(2r)$, $SO(2r+1)$ and $Sp(2r)$ with even $N_f$ matter hypermultiplets
degenerated in pairs. For these groups, the characteristic polynomial is
\be
P(\lambda,u_{2k})= \lambda^{2r} - \sum_{k=1}^r u_{2k} \lambda^{2r-2k} ~,
\label{pe}
\ee
and the low-energy dynamics of the corresponding $\N=2$ super Yang--Mills theory
is described by the hyperelliptic curves \cite{son}
\beqa
y^2 &=& P^2(\lambda,u_{2k}) - 4\Lambda^{4r-2-2N_f}\lambda^{2}
\prod_{j=1}^{N_f}(\lambda^2-m_f^2) \hspace{2cm} ~~~~~~~~~~ SO(2r) ~,\\
y^2 &=& P^2(\lambda,u_{2k}) - 4\Lambda^{4r-4-2N_f}\lambda^{4}
\prod_{j=1}^{N_f}(\lambda^2-m_f^2) ~~~~~~~~~~~~~~~~~~~~~~ SO(2r+1) ~, \\
y^2 &=&(\lambda^2 P(\lambda,u_{2k})+A_0)^2 - 4\Lambda^{4r+4-2N_f}
\prod_{j=1}^{N_f}(\lambda^2-m_f^2) ~~~~~~~~~~~~~~~~ Sp(2r) ~, \label{sp2r}
\eeqa
with $A_0=\Lambda^{2r-N_f+2}\prod_{j=1}^{N_f}m_f$. In order to treat $Sp(2r)$
on equal footing to the other gauge groups, it is convenient to restrict to the
case where two hypermultiplets are massless, which we denote by $Sp(2r)''$. We
can then write the hyperelliptic curve for all these cases as
\be
y^2 = P^2(\lambda,u_{2k}) - 4\Lambda^{4r-2q-2N_f}\lambda^{2q}
\prod_{j=1}^{N_f}(\lambda^2-m_f^2) ~,
\label{curso}
\ee
where $q=1$ for $SO(2r)$, $q=2$ for $SO(2r+1)$ and $q=0$ for $Sp(2r)''$ (in
this last case, $N_f$ accounts for matter hypermultiplets other than the two
mentioned above). These curves have genus $g=2r-1$, where $r$ is the rank of
the gauge group. Then, if we now adjust masses to come in pairs, the curves
take the form
\be
y^2 = P^2(\lambda,u_{2k}) - 4\Lambda^{4r-2q-2N_f}\left(
\lambda^{q}R(\lambda)\right)^2 ~,
\ee
where
\be
R(\lambda)=\prod_{j=1}^{N_f/2}(\lambda^2-m_f^2) ~.
\label{erre}
\ee
Now, similarly to the $SU(N_c)$ case, we define $W\equiv P/(\lambda^q R)$
and the description of the theory is the same as before, with $N=2r-q-N_f$.
The Seiberg--Witten differential is also given by
$dS_{SW}=\lambda\frac{dw}{w}$, and its variation with respect to the moduli
$u_{2k}$ is
\be
\left . \frac{\partial dS_{SW}}{\partial u_{2k}}\right|_{w} =
\frac{\lambda^{2r-2k-q}}{RW'}\frac{dw}{w} = \lambda^{2r-2k}\frac{d\lambda}{y} =
dv^{2k} ~, ~~~~~~~ k = 1,\ldots,r ~.
\ee
From the original space of holomorphic differentials corresponding to the
hyperelliptic curve of genus $g=2r-1$, one really deals with a subspace of
dimension $r$, which is the complex dimension of the quantum moduli space,
generated by those invariant under the reflection $\lambda\to -\lambda$
\cite{son}. This symmetry, of course, also has to be taken into account in the
definition of $d\hat{\Omega}_n$. Among the meromorphic differentials
(\ref{domegan}), only those with odd $n$ are invariant under this reflection.
That is, the differentials we have to consider are
\be
d\hat{\Omega}_{2n-1}\equiv \left( \left[W^{\frac{2n-1}{N}}
\right]_+-\frac{2}{2r-q}\sum_{f=1}^{N_f/2}\left . \left[W^
{\frac{2n-1}{N}}\right]_+ \right|_{\lambda=m_f}
\right)\frac{dw}{w} ~.
\ee
With these remarks in mind, one can proceed along the calculations of the
preceding sections obtaining analogous results. Clearly, these theories have
only odd Whitham times. The recovery of the Seiberg--Witten solution also goes
as in the $SU(N_c)$ case. That is, one must rescale the times $\bar T^{2n-1} =
(T^1)^{-(2n-1)} T^{2n+1}$, the moduli $\bar u_{2k} = (T^1)^{2k}~u_{2k}$, and
the masses $\bar m_{f} = T^1~ m_f$ so, for example, $\alpha^i$ reads
\be
\alpha^i(u_{2k},T^1,\bar T^3,\bar T^5,...) = \sum_{n\geq 1} {\bar T^{2n-1}
\over 2\pi i} \oint_{A^i}{W^{\frac{2n-1}{2r-q-N_f}}_+(\bar u_{2k}) W'(\bar
u_{2k})\ov \sqrt{W^2(\bar u_{2k}) - 4(T^1)^{2N}}} d\lambda ~.
\ee
Then, $\alpha^i(u_{2k},T^1,\bar T^{2n-1>3} = 0) = T^1 a^i(u_{2k},1) =
a^i(\bar u_{2k},\Lambda=T^1)$. In summary, it is clear at this point that the
same formulas (\ref{lasecpri})--(\ref{higherdos}) are obtained for the
first and second derivatives of the effective prepotential, provided the
appropriate value of $W$ and $N$ is considered, and changing $n,m,...$ by
$2n-1,2m-1,...$. Similar checks to the ones discussed in the previous section
were carried out in these cases. Moreover, the semiclassical expansion of the
prepotential up to arbitrary instantonic corrections can be recursively
obtained from these equations in a remarkably simple way \cite{egrm}.

\section{Concluding Remarks}

In this paper we have undertaken the embedding of the Seiberg--Witten ansatz
for the low-energy effective dynamics of $\N=2$ supersymmetric gauge theories
with an even number of massive fundamental hypermultiplets within a Whitham
hierarchy. Aside from its mathematical beauty, this formalism leads to new
differential equations for the effective prepotential that can be easily
applied to obtain powerful results as its whole semiclassical expansion
\cite{egrm}. The expressions obtained by these means are also consistent with
the duality properties of the effective couplings, including those resulting
from the derivation of the prepotential with respect to Whitham times. 

This work opens, or suggests, several interesting avenues for further research.
The most immediate one seems to be its generalizaton to any number of matter
hypermultiplets non-degenerated in mass. This is quite problematic within our
present approach. A possible derivation of the corresponding equations for
arbitrary masses is to use the $u$-plane integral of the topological theory
\cite{mw}. The second derivatives of the prepotential with respect to higher
Whitham times can be understood in that context as contact terms \cite{mm}. At
the same time, these contact terms can be obtained from the behavior under
blowup of the twisted low-energy effective action \cite{losev}. One should be
able to generalize the arguments explained in Ref.\cite{marijapon} to the case
of theories with massive hypermultiplets. The first step in this direction would
be to generalize the $u$-plane integral of \cite{mm} in order to extract the
corresponding blowup formula, and from it one could read the appropriate theta
function. The possible ambiguities in this derivation can be fixed in principle
by looking at the behavior at infinity, as explained in \cite{marijapon}.

The expressions that we have provided extend in most cases the ones in
\cite{ITEP} to higher times, $T^n, n<2N_c$.  There is an obvious question
about the microscopical origin of these deformations. In two dimensional
topological conformal field theory they correspond to marginal deformations
by gravitational descendants. It would be very interesting to have a
clear understanding of the corresponding operators here.

In Ref.\cite{emm}, it was shown that the Whitham times provide generalized
spurionic sources for soft breaking of supersymmetry down to $\N=0$. In this
spirit, the additional times $T^{0,f}$ would also admit an interpretation as
spurion superfields. The formulas we have obtained in the present paper are
ready for use in a study of such softly broken theories along the lines of
\cite{emm,luisIyII}.

Another interesting possibility is the use of the {\em new} equations
(\ref{lasecpri})--(\ref{higherdos}) to study the strong coupling expansion of
the prepotential near the singularities of the quantum moduli space, as it was
done in Ref.\cite{edemas} for the case of pure $SU(N_c)$. Finally, another
avenue for future research is, certainly, the connection of this formalism with
the string theory and D-brane approach to supersymmetric gauge theories, where
some steps has already been given in the last few years \cite{intbrane}. We
believe that all these matters deserve further study.

\section*{Acknowledgements}

We would like to thank Andrei Marshakov and Alexei Morozov for interesting
comments. J.M. wants to thank Christian Klein for discussions and
bibliographical support. The work of J.D.E. has been supported by the Ministry
of Education and Culture of Spain and the National Research Council (CONICET)
of Argentina. The work of M.M. is supported by DOE grant DE-FG02-92ER40704. The
work of J.M. was partially supported by DGCIYT under contract PB96-0960 and
European Union TMR grant ERBFM-RXCT960012.

\appendix
\section{Riemann bilinear relations.}
\setcounter{equation}{0}
\setcounter{subsubsection}{0}

We denote by $\tilde\Gamma$ the cut-Riemann surface, that is the surface with
boundary obtained by removing all $A^i$- and $B_i$-cycles from $\Gamma$. Let
$A^{i\pm}$ and $B_j^{\pm}$ denote the left and right edges of the appropriate
cuts,
\be
\d\tilde\Gamma = \sum_{j=1}^g \left( A^{j+}  +B^+_j -A^{j-} - B_j^- \right)~.
\label{contorno}
\ee
Any abelian differential of the first or second kind is single-valued on
$\tilde\Gamma$. It is sufficient to require that the integration path should
not intersect any $A^i$- or $B_i$-cycles. At the boundary $\d\tilde\Gamma$, the
abelian integral $\Omega(P)$ satisfies
\beqa
\left.\Omega(P)\right\vert_{A^{j+}} -\left.\Omega(P)\right\vert_{A^{j-}} &=& -
\oint_{B_j}d\Omega~~,  \nonumber\\
\left.\Omega(P)\right\vert_{B_j^+} -\left.\Omega(P)\right\vert_{B_j^-} &=&
\oint_{A^j}d\Omega~.
\label{masmenos}
\eeqa
To distinguish a single-valued branch on a third-kind differential,
$d\Omega^{P_a,P_0}_0$, it is necessary to draw additional cuts $\gamma_a$ on
the surface $\tilde\Gamma$ that run from $P_0$ to $P_a$. Let $\gamma_a^\pm$
denote both sides of the cut, then
\be
\left.\Omega(P)\right\vert_{\gamma_a^+} -
\left.\Omega(P)\right\vert_{\gamma_a^-} = 2\pi i ~\res_{P_a}
d\Omega^{P_a,P_0}_0 = - 2\pi i ~\res_{P_0} d\Omega^{P_a,P_0}_0 = 2\pi i ~.
\ee
Most of the manipulations involved in the proofs of the consistency relations
among derivatives of the prepotential rely heavily on the next result: let
$d\Omega$ and $d\Omega'$ be two abelian differentials, then
\be
 {1\ov 2\pi i}\oint_{\d\tilde\Gamma} \Omega d\Omega' =
\sum_{k=1}^g ~~\oint_{A^k}d\Omega \oint_{B_k}d\Omega'  - \oint_{B_k}d\Omega
\oint_{A^k}d\Omega' ~.
\label{laide}
\ee
Notice that this is also true for $d\Omega$ and $d\Omega'$  being, just, closed
differentials. Applying the residue theorem to the left hand side, we obtain
 various relations for the periods of abelian integrals:
\bd
\item{$(i)$}
If  $d\Omega$ and $d\Omega'$ are meromorphic of the first kind (\ie
holomorphic), then
\be
\sum_{k=1}^g ~~\oint_{A^k}d\Omega \oint_{B_k}d\Omega'  - \oint_{B_k}d\Omega
\oint_{A^k}d\Omega'~ = 0 ~.
\label{fRiem}
\ee
In particular, for a  canonical basis of holomorphic
differentials $d\om_i$, $\oint_{A^i}d\om_j = \delta^i{_j}$, we get that
\be
\tau_{ij} = \tau_{ji}~, \label{tausime}
\ee
where $\tau_{ij} = \oint_{B_i}d\om_j$ is the period matrix of
$\Gamma_g$.

\item{$(ii)$}
For $d\Omega=d\om_j$ a holomorphic differential in a canonical basis, and
$d\Omega'=d\Omega^{P_a}_n= (\xi_a^{-n-1}+O(1)) d\xi_a $ a meromorphic
differential of the second kind normalized as in (\ref{merseco}) and
(\ref{mersecod}),  we find
\be
\oint_{B_j} d\Omega^{P_a}_n = {1\ov 2\pi i\,
n}\oint_{P_a}\xi_a^{-n}d\om_j(\xi_a) = ~{1\ov 2\pi i n}\res_{P_a} ( \xi_a^{-n}~
d\om_j(\xi_a) ) ~.
\label{ciclob2}
\ee

\item{$(iii)$}
Again let $d\Omega= d\om_j$ be holomorphic in a canonical basis, and
$d\Omega^{P_a,P_0}_0 $ a meromorphic differential of the third kind, then
\beqa
\oint_{B_j}d\Omega^{P_a,P_0}_0 &=& {1\ov 2\pi i}\oint_{P_a} \om_j ~(\xi_a^{-1}
+{\cal O}(1))d\xi_a - {1\ov 2\pi i} \oint_{P_0} \om_j ~(\xi_0^{-1}+{\cal
O}(1))d\xi_0  \cr &=& {1\ov 2\pi i} \int^{P}_{P_0} d\om_j ~.
\label{hololog}
\eeqa

\item{$(iv)$}
If both $d\Omega =d\Omega^{P_a}_n$ and $d\Omega'=d\Omega^{P_b}_m$ are of the
second or third kind ($m,n=0,1,2,...$), and normalized as in (\ref{merseco})
and (\ref{mersecod}), or (\ref{mert}), then the r.h.s. of (\ref{laide})
vanishes and we obtain the symmetry relations
\beqa
{1\ov n} \res_{P_a}~\xi_{a}^{-n}d\Omega^{P_b}_m &=& {1\ov m}
              \res_{P_b}~\xi_b^{-m}d\Omega^{P_a}_n ~,
\label{emene}\\
{1\ov n}\oint_{P_a} \xi_a^{-n} d\Omega^{P_b,P_0}_0 &=&  \int^{P_a}_{P_0}
d\Omega^{P_a}_n ~.
\label{ceroeme}
\eeqa
\ed

\section{Miscelaneous proofs}
\setcounter{equation}{0}
\setcounter{subsubsection}{0}

Let us show here, for completeness, some of the propositions claimed in this
article.

\noindent
{\em Proof of} ~{\bf Lemma A:}~ Taking into account (\ref{varicurve}) and
(\ref{dif}), we obtain
\beqa
\left. \pder{d\hat\Omega_n}{u_k}\right\vert_w & = & \left(
\d_{u_k}\left[W^{\frac{n}{N}}\right]_+ + \left[W^{\frac{n}{N}}\right]_+'
\d_{u_k}\lambda - \d_{u_k}\kappa_n \right) \frac{dw}{w} \nonumber \\
& = & \sqrt{F} \left( W'\d_{u_k} \left[W^{\frac{n}{N}}\right]_+ -
\left[W^{\frac{n}{N}}\right]_+^{'} \d_{u_k} W - W'\d_{u_k}\kappa_n 
\right) \frac{d\lambda}{y} \label{hu}\\
& = & \left( \left( P'- \med\frac{F'}{F}P\right)\d_{u_k}
\left(\left[W^{\frac{n}{N}}\right]_+ -\kappa_n\right) -
\left[W^{\frac{n}{N}}\right]_+^{'} \d_{u_k} P \right)\frac{d\lambda}{y} ~, 
\eeqa
which exhibits poles at $m_f^\pm$ with residue (\cf Eq.(\ref{rmdn}))
\be
\left.\res_{m_f^\pm} \pder{d\hat\Omega_n}{u_k}\right\vert_w =  \pm
\pder{b_n^f}{u_k} ~.
\ee
To see what is the behaviour at  $\infty^\pm$, we recast (\ref{hu}) in the
following form:
\beqa
\left. \pder{d\hat\Omega_n}{u_k}\right\vert_w & = & \,\sqrt{F} \left(~~
W'\left[\d_{u_k} W^{\frac{n}{N}}\right]_+ - \left[W^{\frac{n}{N}\,'}\right]_+
\d_{u_k} W - W'\d_{u_k}\kappa_n \right) \frac{d\lambda}{y} \nonumber\\
&=&   \sqrt{F} \left(-
W'\left[\d_{u_k} W^{\frac{n}{N}}\right]_- +
\left[W^{\frac{n}{N}'}\right]_- \d_{u_k} W - W'\d_{u_k}\kappa_n 
\right) \frac{d\lambda}{y} ~.
\label{adkj}
\eeqa
The highest order in $\lambda$ of the first two terms is $N_c-2=g-1$, so they
yield holomorphic differentials. Only the last one has a pole with residue
\be
\left.\res_{\infty^\pm} \pder{d\hat\Omega_n}{u_k}\right\vert_w = \mp N
\pder{\kappa_n}{u_k} = \pm\sum_{f=1}^{N_f} \pder{b_n^f}{u_k} ~.
\label{resdudom}
\ee
Altogether, these results imply that we can expand in a canonical basis
\be
\left. \pder{d\hat\Omega_n}{u_k}\right\vert_w = \sum_i
\left(\pder{c^i_n}{u_k}\right) d\om_i + \sum_f \left(\pder{b_n^f}{u_k}\right)
d\Omega_{0,f} ~.
\ee
The coefficients in front of the $d\om_i$ are fixed by first contour 
integrating (\cf Eq.(\ref{aci})) and, afterwards, taking the derivative
$\d_{u_k}$. This is the desired result and a similar analysis can be
performed concerning the derivatives of this object with respect to the
parameters $m_f$. Indeed, we can compute
\beqa
\left. \pder{d\hat\Omega_n}{m_f}\right\vert_w & = & ~\sqrt{F} \left(
W' \d_{m_f}\left[W^{\frac{n}{N}}\right]_+ - \left[W^{\frac{n}{N} }\right]'_+
\d_{m_f} W - W'\d_{m_f}\kappa_n \right) \frac{d\lambda}{y} \label{ju} \\
& = & \left(\left(P'- \med\frac{F'}{F}P\right)
\left(\d_{m_f}\left[W^{\frac{n}{N}}\right]_+(\lambda) -
 \d_{m_f}\kappa_n\right) -  
\frac{\left[W^{\frac{n}{N}}\right]_+ '(\lambda)}{(\lambda-m_f)}
\right)\frac{d\lambda}{y} ~.
\eeqa
Again, there are just poles at $m_g^\pm$ whose residues are 
\beqa
\res_{ m_g^\pm} \left. \pder{d\hat\Omega_n}{m_f}\right\vert_w &=&
\mp\left(\d_{m_f}
\left[W^{\frac{n}{N}}\right]_+ \right) (m_g) \mp\d_{m_f} \kappa_n\pm
\left[W^{\frac{n}{N} }\right]'_+(m_g)\, \delta_{fg} \nonumber\\
&=& \mp \d_{m_f}\left(\left[W^{\frac{n}{N}}\right]_+( \lambda = m_g) -
\kappa_n\right) \nonumber\\
&=&  \pm\pder{b^g_n}{m_f} ~.
\eeqa
At $\infty^{\pm}$, the same trick is in order, namely from (\ref{ju})
one easily gets
\beqa
\left. \pder{d\hat\Omega_n}{m_f}\right\vert_w & = &  \sqrt{F}
\left(-W'\left[\d_{m_f} W^{\frac{n}{N}}\right]_- +
\left[W^{\frac{n}{N}'} \right]_- \d_{m_f} W  - W'\d_{m_f}\kappa_n \right)
\frac{d\lambda}{y} ~,
\eeqa
and, as in (\ref{adkj}), this expression is holomorphic at $\lambda = \infty$
except for the pole in the last term which produces a residue
\be
\left.\res_{\infty^\pm}
\pder{d\hat\Omega_n}{m_f}\right\vert_w =\pm N  \d_{m_f}\kappa_n
 = \mp   \sum_{g=1}^{N_f} \pder{b_n^g}{m_f} ~.
\ee
This proves that the decomposition given in Eq.(\ref{vo}) is correct.

\noindent
{\em Proof of} ~{\bf Lemma B:}~ We want to compute the partial derivative of
the generating meromorphic differential $dS$ with respect to the flat moduli
and Whitham slow times. We have
\beqa
\pder{dS}{T^n} &=& \pder{~}{T^n} \sum_{m\geq 1} T^m d\hat\Omega_m \nonumber\\
&=& d\hat\Omega_n + \sum_{m\geq 1}
T^m \left(\pder{u_k}{T^n}\pder{d\hat\Omega_m}{u_k} +
\pder{m_s}{T^n}\pder{d\hat\Omega_m}{m_s}\right) ~, \label{stepone}
\eeqa
which, after (\ref{va})--(\ref{vo}) and (\ref{vo2})--(\ref{vo3}), reads
\beqa
\pder{dS}{T^n} &=& d\hat\Omega_n + \sum_{m\geq 1} T^m \left(\pder{u_k}{T^n}
\pder{c^i_m}{u_k} d\omega_i + \pder{u_k}{T^n}\pder{b_m^f}{u_k} d\Omega_{0,f}
\right. \nonumber\\ &&~~~~~~~~~~~~~~~~~~~\left. \pder{m_s}{T^n}
\pder{c^i_m}{m_s} d\omega_i + \pder{m_s}{T^n}\pder{b_m^f}{m_s} d\Omega_{0,f}
\right) \nonumber\\ & = & d\hat\Omega_n   - c_n^i d\om_i
- b_n^f d\Omega_{0,f} \nonumber\\
& = & d\Omega_n ~.
\eeqa
Following similar steps, it is straightforward to prove the remaining
propositions, $\pder{dS}{\alpha^i}= d\om_i$ and $\pder{dS}{T^{0,f}} =
d\Omega_{0,f}$.

\section{Meromorphic differentials and the Szeg\"o kernel}
\setcounter{equation}{0}
\setcounter{subsubsection}{0}

In this appendix we shall give the details leading to expressions (\ref{ss})
and (\ref{dod}) in the main text. As we say in Section 5 the residues appearing
in the calculations of the second derivatives of the prepotencial with respect
to $T^n$ and $T^m$ and  with respect to $T^n$ and $T^{0,f}$ involve
differentials defined with respect to $\lambda$ (not to $w$).  This
differentials can be computed using the Szeg\"o kernel, as will be shown in
the next subsections.

\subsection{Second Kind Differentials}

Meromorphic differentials of second kind, $d\tilde\Omega_n(\lambda)$,
are generated by some bi-differential $W(\lambda,\mu)$
upon expanding it around $\mu\to \infty_\pm$.
\be
W(\lambda,\mu)  \stackrel{\mu\to\infty_\pm}{\longrightarrow}   - \sum_{p\geq 1}
d\tilde\Omega^\pm_p (\lambda){d\mu\ov \mu^{p+1}}~.
\label{expanmero}
\ee
The key ingredient is the so-called Szeg\"o kernel
\be
\Psi_e(\lambda,\mu) = \frac{\Theta_e(\vec \lambda- \vec\mu)}{\Theta_e(\vec
0) E(\lambda,\mu)}
\label{sk}
\ee
where $E(\lambda,\mu)$ is the Prime form.
In terms of the Szeg\"o kernel,
\be
W(\lambda,\mu) = \Psi_e(\lambda,\mu)\Psi_{-e}(\lambda,\mu) -
d\om_i(\lambda)d\om_j(\mu)
\left(\frac{1}{i\pi} \pder{~}{ \tau_{ij} }\log \Theta_E(\vec0)\right).\label{c1}
\ee
$\Psi_e(\lambda,\mu) $ is a $1/2$-bidifferential that has a simple
hyperelliptic representation
whenever $e$ denotes an even non-singular characteristic $e=-e= E$.
 Such characteristics are in one-to-one correspondence
with the partitions of the set of $2g+2$ ramification points into two equal
subsets,
$ \{ r_\alpha^\pm ~,\alpha = 1,2,...,g+1 \} $,
such that $y^2(\lambda) = \prod_{\alpha=1}^{g+1}(\lambda -
r_\alpha^+)(\lambda - r_\alpha^-) =
Q_+(\lambda)Q_-(\lambda)$.
In this particular case \cite{fay}
\be
\Psi_E(\lambda,\mu) = \frac{U_E(\lambda) + U_E(\mu)}{2\sqrt{U_E(\lambda)
U_E(\mu)}}
\frac{\sqrt{d\lambda d\mu}}{\lambda-\mu}
\label{sze}
\ee
where
\be
U_E(\lambda) = \sqrt{\prod_{\alpha=1}^{g+1}
\frac{\lambda-r_\alpha^+}{\lambda-r_\alpha^-}} =
\frac{1}{y(\lambda)}\prod_{\alpha=1}^{g+1} (\lambda-r_\alpha^+) =
\frac{Q_+(\lambda)}{ y(\lambda)}~. \label{uuuu}
\ee
An explicit calculation yields
\be
\Psi_E^2(\lambda,\mu) =
\frac{ Q_+(\lambda)Q_-(\mu) + Q_+(\mu)Q_-(\lambda) + 2
y(\lambda)y(\mu)}{4y(\lambda)y(\mu)}
\frac{d\lambda d\mu}{(\lambda-\mu)^2}~.
\ee

If we now assume the symmetric scenario, \ie values of $m_f$ come in pairs,
then there is as privilegiate choice for $Q_\pm = P \pm 2\sqrt{F}$. The
characteristic $E$ appearing in (\ref{c1}) is associated with the splitting of
the roots of the discriminant \cite{rauch}\cite{ITEP} and for this particular
case we have that \cite{emm}
\be
E = \left[ {\vec\alpha\atop \vec\beta}\right] ~~~~~\longrightarrow~~~~~
\vec \alpha=(0, \dots, 0) ~~~\hbox{and}~~~ \vec \beta = (1/2, \dots, 1/2),
\label{carac}
\ee
and the theta function involved in the above equations is then
\be
 \Theta_{E=[\vec \alpha, \vec \beta]}(\vec{z} \vert \tau)=\sum_{\vec{n}\in
Z } e^{i \pi \tau_{ij} n_i n_j +  2 \pi i \sum_k(z_k+\frac12)n_k},
\label{theta}
\ee
so we have in (\ref{c1}) that $\frac{1}{(2\pi
i)^2}\partial^2_{i,j}\log\Theta_E=\frac{1}{\pi
i}\partial_{\tau_{ij}}\log\Theta_E$.

For this particular choice of $Q_\pm$
\beqa
\Psi_E^2(\lambda,\mu) &=&
\frac{ P(\lambda)P(\mu) -4\sqrt{F(\lambda)}\sqrt{F(\mu)}+ 
y(\lambda)y(\mu)}{2y(\lambda)y(\mu)}
\frac{d\lambda d\mu}{(\lambda-\mu)^2} \nonumber\\
&\stackrel{\mu\to\infty_\pm}{\longrightarrow}& \sum_{p=1}^{2N}
\left(\frac{\pm P(\lambda)+y(\lambda)}{2y(\lambda)}\lambda^{p-1}d\lambda
\right. \nonumber\\ 
&& \left.
~~~\pm\theta(p-N-1)\sum_{k=0}^{p-N-1}c_k\frac{2(N+k-p)}{p}
\frac{\sqrt{F(\lambda)}}{y(\lambda)}\lambda^{p-k-N-1}d\lambda\right)
\frac{pd\mu}{\mu^{p+1}}
\nonumber\\
&& ~~~~~~~+ ~ {\cal O}(\mu^{-2N-2}) \label{cenu}
\eeqa
where we have expanded $W^{-1}(\mu) = \sum_{k=0}^\infty c_k/\mu^{k+N}$.
Next, expand  $d\om_j(\mu)$ around $\infty^\pm$  
\be
d\om_j(\mu)\stackrel{\mu\to\infty_\pm}{\longrightarrow}
\pm\sum_{p=1}^{2N-1}\left(\frac{1}{p}\pder{h_{p+1}}{a^j}\right)
\,\frac{pd\mu}{\mu^{p+1}} ~+ ~{\cal O}(\mu^{-2N-1})\label{cedi}
\ee
With  this setup in mind, comparing with (\ref{cenu}) and (\ref{cedi})
with (\ref{expanmero}) we obtain   for $p<2N$
\beqa
d\tilde\Omega_{p}(\lambda)&=&d\tilde\Omega_{p}^{+}(\lambda)-
d\tilde\Omega_{p}^{-}(\lambda)\nonumber\\
&=&
-\lambda^{p-1}\frac{P(\lambda)}{y(\lambda)}d\lambda
+d\om_i(\lambda)\,\frac{2}{p}\frac{\partial h_{p+1}}
{\partial a_j}\frac{1}{\pi
i}\partial_{\tau_{ij}}\log\Theta_E \nonumber\\
&& ~~ -  ~     \theta(p-N-1)\sum_{k=0}^{p-N-1}c_k\frac{4(N+k-p)}{p}
\frac{\sqrt{F(\lambda)}}{y(\lambda)}\lambda^{p-k-N-1}d\lambda ~.
 \label{c12}
\eeqa

\subsection{Third kind differential}
The third kind meromorphic differential $d\Omega_0^{P,Q}(\lambda)$ with
vanishing $A^i$-cycles can be written in terms of the Prime form \cite{fay,mum}
as follows
\be
d\Omega_0^{P,Q}(\lambda) = d\log\frac{E(\lambda,P)}{E(\lambda,Q)} ~.
\label{thprime}
\ee
Also, an explicit representation in terms of the Szeg\"o kernel (\ref{sze}) can
be found (see Proposition 2.10 in Ref.\cite{fay}),
\be
d\Omega_0^{P,Q}(\lambda) =
\frac{\Psi_e(\lambda,P)\Psi_{-e}(\lambda,Q)}{\Psi_e(P,Q)} - 
d\om_i(\lambda){1\ov 2\pi
i}\pder{~}{z^i}\left[\log\Theta_e(\vec{z}_{P,Q} |\tau)
-\log\Theta_e(   0  |\tau)\right] ~,
\label{tkd}
\ee
where
$$ \vec{z}_{P,Q} = \frac{1}{2\pi i} \int^Q_Pd\vec\om ~, $$
is the image of the divisor $Q-P$ under the Abel map.
When $e = -e = E$ is an even half-integer characteristic,
$\d_i\log\Theta(\vec{z}\mid\tau)$ is odd under $\vec z\to -\vec z$ so 
$\d_i \log \Theta_E(0|\tau) = 0$. Letting $P = m_f^\pm$ and $Q =
\infty^\pm$, and making use of (\ref{sze}) and (\ref{uuuu}) as well as
  $U(\infty^{\pm})= U(m_f^\pm)=\pm1$ the third
kind differential reads
\be 
d\Omega_0^{m_f^\pm,\infty^\pm}(\lambda) = \pm \frac{(U_E(\lambda) \pm
1)^2}{4U_E(\lambda)} \frac{d\lambda}{\lambda - m_f} ~\pm~ d\om_i(\lambda){1\ov
2\pi i}\pder{~}{z^i}
\log\Theta_E(\vec{z}_f|\tau) ~, 
\label{fco}
\ee 
where $\vec{z}_f$ is given in Eq.(\ref{abelpoint}). This third kind
differential is easily seen to have simple poles at $m_f^\pm$ with residue
$+1$ and $\infty^\pm$ with residue $-1$, while being regular everywhere else.
Hence, we find that $d\Omega_{0,f} = d\Omega_{0}^{m_f^+,\infty^+} -
d\Omega_{0}^{m_f^-,\infty^-}$ can be written as 
\be
d\Omega_{0,f}(\lambda) =  \frac{P}{y} \frac{d\lambda}{\lambda - m_f} ~+~ 
d\om_j(\lambda){1\ov
 \pi i}\pder{~}{z^j} \log\Theta_E(\vec{z}_f|\tau) ~,
\label{c16}
\ee
which is the result we need to evaluate the second derivative of the
prepotential $\F$ with respect to $T^n$ and $T^{0,f}$.


\end{document}